\documentclass[final,3p,twocolumn]{elsarticle}
\biboptions{sort&compress}

\usepackage{lineno,hyperref}
	\hypersetup{colorlinks=true,pdfborder=0 0 0}
\usepackage{subcaption}
\usepackage{amsmath,siunitx,multirow,multicol}
\usepackage{chemformula}
\usepackage{amssymb}
\usepackage{gensymb}
\usepackage{cleveref}
\usepackage{siunitx}
\usepackage[dvipsnames]{xcolor}
\modulolinenumbers[1]
\usepackage{graphicx}
\usepackage{float}
\usepackage{fontspec}
\usepackage{xspace}

\usepackage{fancyhdr}

\usepackage{etoolbox}
\makeatletter
\patchcmd{\ps@pprintTitle}
  {Preprint submitted}
  {Postprint of article submitted}
  {}{}
\makeatother

\journal{Calphad}

\begin{document}

\begin{frontmatter}

\title{Thermodynamic assessment of the Ba--La--S and Ga--La--S systems}

\address[az-mse]{Department of Materials Science and Engineering, University of Arizona, Tucson, AZ 85721, USA}
\address[az-am]{Graduate Interdisciplinary Program in Applied Mathematics, University of Arizona, Tucson, AZ 85721, USA}

\author[az-mse]{Jiayang Wang}
\author[az-mse]{Guangyu Hu}
\author[az-mse]{Pierre Lucas}
\author[az-mse,az-am]{Marat I. Latypov\corref{cor1}}
\cortext[cor1]{corresponding author}
\ead{latmarat@arizona.edu}

\begin{abstract}
    
This paper presents the first thermodynamic assessment of binary and pseudo-binary phase diagrams in the Ba--La--S and Ga--La--S systems by means of the CALPHAD method. Experimental phase diagram equilibrium data and thermodynamic properties available from the literature were critically reviewed and assessed using thermodynamic models for the Gibbs energies of individual phases. The associated solution model was used to describe the short-range ordering behavior of the liquid phases. To supplement the limited experimental data reported in the literature, ab initio molecular dynamics calculations were employed to derive mixing enthalpies of the liquid phases in the binary subsystems. The resulting phase diagrams and calculated thermodynamic properties show good agreement with available literature within the investigated compositional ranges of binary and pseudo-binary systems. 

\noindent Note: \textcolor{red}{this is an author-generated postprint} of the article by Wang et al.\ \href{https://doi.org/10.1016/j.calphad.2026.102924}{published} in {\it{Calphad}} (2026). \\ DOI for version of record: 10.1016/j.calphad.2026.102924

\begin{keyword}
Chalcogenides  \sep phase diagrams \sep thermodynamic assessment \sep CALPHAD \sep Ab-initio calculations
\end{keyword}

\end{abstract}

\end{frontmatter}

\pagestyle{fancy}
\fancyhf{}
\fancyhead[LO]{Postprint of \href{https://doi.org/10.1016/j.calphad.2026.102924}{Wang et al., Calphad 92 (2026) 102924}}

\section{Introduction}

Refractory rare-earth sulfide materials are of increasing interest for applications in  infrared (IR) optics, semiconductor devices, and protective coatings \cite{white1990refractory, mccloy2013infrared} thanks to their excellent combination of mechanical, thermal, and optical properties \cite{mccloy2013infrared, mccloy2015infrared}. For example, high melting points of such sulfide compounds as \ch{BaS} ($T_m$ = \SI{2454}{K}) \cite{boury2021liquid}, \ch{La2S3} ($T_m$ = \SI{2211}{K}) \cite{kamarzin1981growth}, and \ch{Ga2S3} ($T_m$ = \SI{1393}{K}) \cite{massalski1990binary} make them promising candidates for high temperature applications. Nevertheless, rare-earth sulfides remain an emerging materials class with significantly fewer studies to date compared to more established systems such as oxide-based materials \cite{kumta1994rare}. 

Supporting further development and deployment of rare-earth sulfides requires a deep understanding of their phase stability and thermodynamic behavior. Most of the published knowledge on phase equilibria and thermodynamic properties of these materials is based on direct experimental observations with a noticeable lack of thermodynamic models developed with the modern computational approaches, such as the CALPHAD (CALculation of PHAse Diagrams) method \cite{hallstedt2025sgte}. Reliance solely on experiments hinders efficient exploration of the compositional design space in pursuit of advanced performance. In this context, the development of computational thermodynamic models are essential for guiding targeted synthesis and processing as well as for predicting properties without extensive experimentation.

This work aims to fill this gap for two sulfide systems: La--Ba--S and La--Ga--S, both of which have strong potential for advanced applications offering high thermal stability with broad IR transparency and glass-forming ability \cite{petracovschi2014synthesis, loireau1977systemes}. Specifically, we establish a thermodynamic database that contains thermodynamic models for all experimentally observed phases calibrated to experimental and computational data available in the literature. 
We further supplemented literature data with thermodynamic property data from ab initio molecular dynamics (AIMD) simulations of liquid phases, for which published information is very limited.
To this end, we first review the literature for the relevant binary and ternary systems and summarize the phases considered for the thermodynamic assessment in this study (\Cref{sec:lit}). We then describe the thermodynamic models adopted for the phases, fit the corresponding parameters to the compiled literature data (\Cref{sec:models}) and our AIMD results (\Cref{sec:aimd_results}). Finally, we compare the calculated phase diagrams and thermodynamic properties with both experimental and simulation data for these systems (\Cref{sec:assessment}). Given a limited scope of published data on the systems of interest, we focus on binary and pseudo-binary phase diagrams in this study. 

\section{Literature review}
\label{sec:lit}

The CALPHAD method is a hierarchical and self-consistent framework for developing thermodynamic models of multicomponent systems \cite{lukas2007computational}. In line with the CALPHAD approach, this section reviews experimental data that are directly relevant to the two systems examined in this study. Specifically, we cover all the binary systems (Ba--S, Ba--La, La--S, Ga--S) except Ga--La, for which a thermodynamic description is already available from Boudra et al.~\cite{boudraa2022thermodynamic}. \Cref{tab:Ba-La-S_phases,tab:Ga-La-S_phases} summarize all the phases (and their crystal structures) in the Ba--La--S and Ga--La--S systems considered in this study as a result of the literature review detailed below. 

\subsection{Phase diagram data}

{\underline{Barium--Sulfur}}. The Ba--S phase diagram has been reported for a limited composition range between 50 and 80~at.\SI{}{\percent} of sulfur by Robinson et al.\ \cite{robinson1931xcv}, who used thermal analysis. This restriction arises because, at the high temperatures required to investigate Ba-rich and S-rich alloys, containing sulfur vapor in sealed vessels is extremely challenging: sulfur is highly reactive with refractory container materials, making it difficult to maintain a fixed overall composition during equilibration and thus limiting the experimentally accessible portion of the diagram \cite{jak2008phase, feutelais2002standard}. Within this accessible range, Robinson et al.\ identified several invariant reactions by thermal arrest measurements, including the melting temperature of high-temperature phase ht-\ch{BaS2} (\SI{1198}{K}), the phase transition from ht-\ch{BaS2} to lt-\ch{BaS2} at \SI{973}{K}, and the melting temperature of \ch{BaS3} (\SI{827}{K}) \cite{robinson1931xcv}. 
The ht-\ch{BaS2} to lt-\ch{BaS2} transition was later confirmed by Kawada et al.\ \cite{kawada1975barium}. For \ch{BaS}, early structural studies date back to 1923, when Holgersson \cite{holgersson1923struktur} determined its crystal structure by X-ray diffraction (XRD), later refined by Güntert and Faessler \cite{guntert1956prazisionsbestimmung}. More recent work reported the melting point of \ch{BaS} as \SI{2454}{K} \cite{boury2021liquid}, \SI{2470}{K} \cite{livey1959high}, and \SI{2480}{K} \cite{holcombe1984tentative}. Because \ch{BaS} is highly volatile near \SI{2500}{K}, small differences in sulfur vapor pressure control, sample purity, and measurement method vary the reported melting point. Metastable \ch{Ba2S3} (studied by Yamaoka et al.\ \cite{yamaoka1975structural}) forms only under high-pressure conditions and is absent from the phase diagram at ambient pressure. Regarding the liquidus, Robinson et al.\ \cite{robinson1931xcv} reported only a single data point between \ch{BaS} and \ch{BaS2}. Between \ch{BaS2} and \ch{BaS3}, they observed a sharp liquidus drop of \SI{350}{\celsius} for just a \SI{4}{\percent} change in composition, whereas, beyond \ch{BaS3}, the liquidus line flattens considerably. As liquidus data are scarce, it remains unclear whether the melting of ht-\ch{BaS2} and \ch{BaS3} is congruent or incongruent. Robinson et al.\ speculated a eutectic reaction based on extrapolation. In this work, we fit the liquidus line accordingly and model eutectic melting behavior of the compounds.

{\underline{Barium--Lanthanum}}. The Ba--La phase diagram has received little attention, with the only experimental investigation conducted by Pyagai et al.\ \cite{pegaj1985phase}. Their phase diagram indicates solid solutions in both Ba- and La-rich composition ranges, with about 0.2~at.\SI{}{\percent} solubility of La in Ba and 1~at.\SI{}{\percent} of Ba in La. The limited solubility indicates positive enthalpy of mixing between Ba and La atoms in the solid phases (b.c.c.-Ba, b.c.c.-La and f.c.c.-La). Below the melting point, the phase diagram is dominated by two-phase regions with no intermetallic compounds. At elevated temperatures above the melting point, a large miscibility gap spans a broad compositional range from approximately 0.4 to 90~at.\SI{}{\percent} of La.

{\underline{Lanthanum--Sulfur}}. The La--S phase diagram was experimentally investigated by Kamarzin et al.\ \cite{kamarzin1981growth} and Mironov et al.\ \cite{mironov1978phase} through visual polythermal method and high-temperature annealing. Based on these studies, Massalski et al.\  \cite{massalski1990binary} assessed and constructed a La--S phase diagram for concentrations of sulfur below 70~at.\SI{}{\percent}. At higher sulfur contents, the vapor pressure of sulfur becomes too high to maintain a well-defined bulk composition during equilibration, preventing reliable phase-diagram measurements in a manner analogous to the Ba--S system. The La--S binary system contains the \ch{LaS} single phase and \ch{La2S3} polymorphic phases. Westerholt et al.\ \cite{westerholt1986magnetic} experimentally determined the crystal structure and lattice parameter of \ch{LaS}, while other researchers measured its congruent melting point as approximately  \SI{2573}{K} \cite{kamarzin1981growth, west1999basic}. For \ch{La2S3}, three polymorphs ($\alpha$, $\beta$, and $\gamma$) have been reported. The polymorph $\alpha$-\ch{La2S3} adopts a \ch{Gd2S3}-type orthorhombic structure and belongs to the $Pnam$ space group \cite{prewitt1968structure}. The $\alpha$ polymorph is stable at low temperatures and undergoes an irreversible transformation to $\beta$-\ch{La2S3} at $650\pm$\SI{50}{\degreeCelsius} \cite{picon1956sous}. The tetragonal $\beta$-\ch{La2S3} with the $I4_{1}/acd$ space group is stabilized as an oxysulfide that replaces some sulfur atoms with oxygen, forming compositions  \ch{La10S_{15-x}O_x},  $0\leq x\leq 1$ \cite{schleid1991m10s14o}. In this structure, each La atom is surrounded by eight S atoms in a dodecahedral arrangement with dodecahedra sharing triangular faces or edges with neighboring units. The transformation from $\beta$-\ch{La2S3} to $\gamma$-\ch{La2S3} occurs at temperatures up to \SI{1300}{\degreeCelsius} \cite{eyring2002handbook}. The $\gamma$-\ch{La2S3} polymorph is a solid solution adopting a cubic \ch{Th3P4}-type structure in the $I\bar{4}3d$ space group with a stoichiometric composition \ch{La_{3-x}$Va$_xS4} $(0 \leq x \leq 1/3)$, where $Va$ denotes vacancies in the lanthanum sublattice, spanning compositions between \ch{La2S3} and \ch{La3S4}. The melting point decreases with composition from \ch{La3S4} to \ch{La2S3} \cite{kamarzin1981growth}. The melting point of \ch{La3S4} was reported as \SI{2383}{K} \cite{kamarzin1981growth}, while  \ch{La2S3} has been reported (by multiple groups \cite{kamarzin1981growth, nikolaev2008vapor, vasilyeva2008high}) to melt in a range between \SI{2133}{} and \SI{2211}{K}. The potential reactivity with crucibles might have affected the accuracy of the results \cite{boury2021liquid}. Kamarzin et al.\ additionally include two eutectic points in the phase diagram: one between \ch{LaS} and \ch{La2S3} at \SI{1985}{K} at 54.6~at.\SI{}{\percent} of sulfur, another one at \SI{1191}{K} and 0.01~at.\SI{}{\percent} of sulfur \cite{kamarzin1981growth}. \ch{LaS2} polymorphs were reported by Dugue et al.\ (orthorhombic, $Pna2_1$) and Le Rolland et al.\ (monoclinic, $P12_1/c1$) \cite{rolland1991vibrational, dugue1978structure}, with a melting temperature of \SI{1917}{K}. 

\begin{table*}[h]
    \centering
    \begin{tabular}{lccccc}
    \hline
        Phase & Prototype & Pearson & Space & Description \\
              &           & symbol  & group &              \\
    \hline
        b.c.c-Ba & Ba & cI2 & $Im3\bar{m}$ & Solid soln.\ in b.c.c Ba & \\
        d.h.c.p-La & $\alpha$-La & hP4 & $P6_3/mmc$ & d.h.c.p La & \\
        f.c.c-La & W & cI2 & $Im\bar{3}m$ & Solid soln.\ in f.c.c La & \\
        b.c.c-La & Cu & cF4 & $Fm\bar{3}m$ & Solid soln.\ in b.c.c La & \\
        S  & $\alpha$-S & oF128 & $Fddd$ & - & \\
        \ch{BaS} & \ch{NaCl} & cF8 & $Fm3\bar{m}$ & Binary phase \ch{BaS} \\
        lt-\ch{BaS2} & \ch{BaS2} & mS16 & $C2/c$ & Binary phase \ch{BaS2} \\
        ht-\ch{BaS2} & \ch{CaC2} & tI6 & $I4/mmm$ & Binary phase \ch{BaS2} \\
        \ch{BaS3} & \ch{BaS3} & tP8 & $P4_21m$ & Binary phase \ch{BaS3} \\
        \ch{LaS} & \ch{NaCl} & cF8 & $Fm3\bar{m}$ & Binary phase \ch{LaS} \\
        $\alpha$-\ch{La2S3} & \ch{Gd2S3} & oP40 & $Pnam$ & Binary phase \ch{La2S3} \\
        $\beta$-\ch{La2S3} & \ch{Ca2Si3} & tI80 & $I4_{1}/acd$ & Binary phase \ch{La2S3} \\
        $\gamma$-\ch{La2S3} & \ch{Th3P4} & cI28 & $I\bar{4}3d$ & Solid soln.\ in cubic \ch{La2S3} \\
        \ch{LaS2} & - & - & - & Binary phase \ch{LaS2} \\
    \hline

    \end{tabular}
    \caption{Phases in the Ba--La--S system considered in this study and their symbols. }
    \label{tab:Ba-La-S_phases}
\end{table*}

{\underline{Gallium--Sulfur}}. The experimental phase diagram of the Ga--S system remains unsettled. The discrepancies among the reported phase diagrams arise from experimental challenges. Specifically, the low volatility of gallium chalcogenides at sulfur mole fractions below 0.6 complicates vapor pressure measurements and impedes the growth of single crystals with uniform composition. The ambiguity of electrical measurements further contributes to inconsistencies in the reported data \cite{zavrazhnov2001manometric}. Below, we outline the two prevailing interpretations of the phase diagram found in the literature.

Rustanov et al.\ \cite{rustanov1967investigation} investigated the system in the sulfur composition range up to 60~at.\SI{}{\percent}, and their findings were summarized in a phase diagram by Massalski et al.\ \cite{massalski1990binary}. This phase diagram includes a narrow liquid--liquid miscibility gap between 2 and 17~at.\SI{}{\percent} sulfur at \SI{1208}{K} as well as solid phases \ch{GaS}, \ch{Ga2S}, \ch{Ga2S3}, and \ch{Ga4S5}. The phases \ch{GaS} and \ch{Ga2S3} have been confirmed by subsequent studies \cite{pardo1993diagramme, zlomanov1987ptx, greenberg2001thermodynamic}, whereas the phases \ch{Ga2S} and \ch{Ga4S5} described by Rustanov et al.\ as incongruently melting intermediate compounds are less established. Although \ch{Ga4S5} was independently observed \cite{spandau1958thermische}, neither the existence nor the structure of \ch{Ga2S} or \ch{Ga4S5} has been confirmed by other reports \cite{zavrazhnov2018phase}. Given these concerns, we do  not consider the intermediate phases \ch{Ga2S} and \ch{Ga4S5} nor the liquid--liquid miscibility gap in the Ga-rich region proposed by Rustanov et al.\ \cite{rustanov1967investigation}.

Several studies proposed an alternative view of the Ga--S phase diagram for the full composition range \cite{pardo1993diagramme, lieth1966p, zlomanov1987ptx, greenberg2001thermodynamic}, which includes two distinct liquid--liquid miscibility gaps: one between \ch{Ga} and \ch{GaS}, and another between \ch{Ga2S3} and \ch{S}. The Ga-rich miscibility gap at \SI{1203}{K} spans a wide range from 10.3 to 48.4~at.\SI{}{\percent} sulfur, a finding supported earlier by Spandau and Klanberg \cite{spandau1958thermische}. Lieth et al. \cite{lieth1966p} also noted that analogous immiscibility regions exist in similar systems such as Ga--Te \cite{newman1961phase}, In--Te \cite{grochowski1964phase}, and In--S \cite{hansen1958co}, which suggests that the monotectic point on the Ga-rich side lies very close to pure Ga. The S-rich liquid--liquid miscibility gap was reported to extend from 70 to 95~at.\SI{}{\percent} sulfur at \SI{1266}{K} \cite{pardo1993diagramme}. For the generally accepted phases \ch{GaS} and \ch{Ga2S3}, Pardo et al.\ \cite{pardo1993diagramme} and Lieth et al.\ \cite{lieth1966p} reported consistent melting points of \SI{1235}{K} for \ch{GaS} and \SI{1363}{K} for \ch{Ga2S3}, both of which are lower than those reported by Rustanov et al.\ \cite{rustanov1967investigation}. This study adopts the phase diagram of Pardo et al.\ and Lieth et al.\ owing to their better consistency and broader acceptance in the literature.

For \ch{GaS} and \ch{Ga2S3}, Zlomanov and Novoselova \cite{zlomanov1987ptx} suggested these phases represent primary stable compounds that form a eutectic system with a melting point above \SI{1223}{K}. More recent studies by Zavrazhnov et al.\ \cite{zavrazhnov2018phase} and Berezin et al.\ \cite{berezin2017фазовая} investigated the high-temperature phase diagram in the 48 to 60.7~at.\SI{}{\percent} sulfur range using differential thermal analysis (DTA) and chromatic--thermal analysis (CrTA). Their results showed that the system consists only of \ch{GaS} and \ch{Ga2S3} at temperatures below \SI{1151}{K}. A higher temperatures (\SI{1151}{} to \SI{1383}{K}), the phase diagram becomes more complex with three distinct phases in a narrow compositional range (59 to 60.7~at.\SI{}{\percent} sulfur): $\sigma$ (\ch{Ga_{0.41}S_{0.59}}), \ch{Ga2S3}$^\prime$, and $\gamma$-\ch{Ga2S3} \cite{zavrazhnov2018phase}. Although the structural details of the $\sigma$ phase are unknown, it is a line compound with about 59~at.\SI{}{\percent} S. This phase is stable only within a limited temperature range (\SI{1151}{} to \SI{1195}{K}) and decomposes at \SI{1195}{K} via a peritectic reaction $\sigma \rightleftharpoons L + \text{Ga}_2\text{S}_3^\prime$. Two polymorphs of gallium sesquisulfide were also identified: a low-temperature monoclinic form ($\gamma$-\ch{Ga2S3}) and a high-temperature wurtzite-type form (\ch{Ga2S3}$^\prime$) \cite{zavrazhnov2018phase}. These phases exhibit nearly identical compositions and coexist at elevated temperatures (near \SI{1243}{K}), suggesting that \ch{Ga2S3}$^\prime$ behaves as a solid solution. The following eutectic reaction occurs at \SI{1184}{K} and 59.5~at.\SI{}{\percent} sulfur: $\sigma + \gamma$-$\text{Ga}_2\text{S}_3 \rightleftharpoons \text{Ga}_2\text{S}_3^\prime$. In addition, $\gamma$-\ch{Ga2S3} undergoes a peritectic decomposition at \SI{1279}{K}:  $\gamma$-$\text{Ga}_2\text{S}_3 \rightleftharpoons L + \text{Ga}_2\text{S}_3^\prime$ \cite{zavrazhnov2018phase}. At higher temperatures, \ch{Ga2S3}$^\prime$ adopts a defective wurtzite structure in which sulfur atoms form a hexagonal close-packed arrangement. Gallium atoms occupy only two-thirds of the upward-oriented $X_4$ tetrahedral sites in an ordered manner. The presence of unoccupied sites in \ch{Ga2S3}$^\prime$ provides the structural basis for its solid-solution behavior  \cite{shimomura2024first, krebs1993synthese}.

Investigations of \ch{Ga2S3} below \SI{1073}{K} revealed three polymorphs: $\alpha$, $\beta$, and $\gamma$ \cite{massalski1990binary,hahn1949ueber, goodyear1963crystal, jones2001refinement}. The low-temperature $\alpha$-\ch{Ga2S3} phase adopts a monoclinic crystal structure in the $Cc$ space group with ordered vacancies \cite{hahn1949ueber, goodyear1963crystal, jones2001refinement}. The $\beta$-\ch{Ga2S3} phase is stable at intermediate temperatures \cite{massalski1990binary}, while the $\gamma$-\ch{Ga2S3} phase is observed at higher temperatures according to the phase diagram proposed by Massalski et al.\ \cite{massalski1990binary}. However, the precise transition temperatures among the \ch{Ga2S3} polymorphs remain uncertain.

{\underline{Barium--Lanthanum--Sulfur}}. Experimental data for the ternary Ba--La--S phase diagram are rather limited. Andreev et al.\ \cite{andreev1991interaction} conducted comprehensive research on \ch{BaS}--\ch{Ln2S3}, including \ch{BaS}--\ch{La2S3} (Ln for Lanthanides). Using XRD, differential thermal analysis (DTA), and visual polythermal analysis, these authors found that the light lanthanide sulfides do not form intermediate compounds but have extensive solid solubilities of \ch{BaS}. The solubility of \ch{BaS} in \ch{Ln2S3} decreases as the atomic number of the lanthanide element increases. Recently, Boury and Allanore \cite{boury2021liquid} investigated the pseudo-binary \ch{BaS}--\ch{La2S3} system above \SI{1573}{K} using high-resolution in situ imaging combined with thermal arrest measurements. The observation of melts in a container-free environment resulted in melting points of the \ch{BaS} and \ch{La2S3} phases equal to \SI{2454}{} and \SI{2004}{K}, respectively. The researchers reported two eutectic points at \SI{1790}{K} and \SI{1805}{K}. However, such in situ imaging captures only the surface phenomena, which may not represent the bulk behavior, especially in the absence of support from direct characterization for phase identification. For this reason,  we exclude these data \cite{boury2021liquid} from further analysis in this study. 

\begin{table*}
    \centering
    \begin{tabular}{lccccc}
    \hline
        Phase & Prototype & Pearson & Space & Description \\
              &           & symbol  & group &             \\
    \hline
        Ga & Ga & oS8 & $Cmca$ & Orthorhombic Ga & \\
        \ch{GaS} & \ch{GaS} & hP8 & $P6_3/mmc$ & Binary phase \ch{GaS} \\
        $\alpha$-\ch{Ga2S3} & wurtzite & mP80 & $Cc$ & Binary phase $\alpha$-\ch{Ga2S3} \\
        $\beta$-\ch{Ga2S3} & $\beta$-wurtzite & hP4 & $P6_3mc$ & Binary phase $\beta$-\ch{Ga2S3} \\
        $\gamma$-\ch{Ga2S3} & sphalerite & cF40 & $F\bar{4}3m$ & Binary phase $\gamma$-\ch{Ga2S3} \\
        
        $\sigma$ & - & - & - & Binary phase $\sigma$ \\
        \ch{Ga2S3}$^\prime$ & wurtzite & hP24 & $P6_1$ & Solid soln.\ in \ch{Ga2S3}$^\prime$ \\

        \ch{LaGaS3} & \ch{La2S3} & mP12 & $P2_1c$ & Ternary phase \ch{LaGaS3} \\
        \ch{La9Ga5S21} &  & hR30 & $R3$ & Ternary phase \ch{La9Ga5S21} \\
    \hline

    \end{tabular}
    \caption{Phases in the Ga--La--S system considered in this study and their symbols. Phases for the Ga--La system, thermodynamically assessed elsewhere \cite{boudraa2022thermodynamic}, are not shown.} 
    \label{tab:Ga-La-S_phases}
\end{table*}

\underline{Gallium--Lanthanum--Sulfur}. Most studies of the Ga--La--S system, focused on the pseudo-binary \ch{Ga2S3}--\ch{La2S3} section. Loireau et al.\ \cite{loireau1977systemes} measured solidus lines and invariant points, and identified two intermetallic compounds: \ch{LaGaS3}, which melts incongruently at \SI{1225}{K}, and \ch{La9Ga5S21}, which melts congruently at \SI{1425}{K}. Bakhtiyarly et al.\ \cite{bakhtiyarly2016ternary} examined the \ch{La2S3}--\ch{Ga2S3}--\ch{EuS} ternary system using differential thermal analysis and XRD. Their results within the \ch{Ga2S3}--\ch{La2S3} subsystem confirmed several invariant reactions identified Loireau et al.\ \cite{loireau1977systemes}. Since no liquidus line was reported for this pseudo-binary system, we fit the liquidus line to invariant reactions, adopting the experimental phase diagram of Loireau et al.\ \cite{loireau1977systemes}.

\subsection{Thermodynamic data}

\subsubsection{Ba--S system}

For the Ba--S system, most available studies focused on \ch{BaS}, whose thermodynamic properties are summarized in the NIST-JANAF Thermochemical Tables \cite{chase1998nist}, including heat capacity $C_p$, the Gibbs free energy of formation $\Delta G$, the enthalpy increment $H(T) - H(298~\mathrm{K})$ and formation enthalpy $\Delta H(298K)$. King and Weller \cite{king1960low} studied heat capacity of \ch{BaS} at low temperatures and reported a room-temperature value of \SI{49.374}{J/mol.K}. Above \SI{300}{K}, heat capacity values were extrapolated using graphical methods and Kubaschewski's method B \cite{kubaschewski1977metallurgical}. Tuncel et al.~\cite{tuncel2009first} computed the thermodynamic properties of \ch{BaS} using the density functional theory (DFT). The DFT results were confirmed by Rino et al.~\cite{rino2014interaction} using molecular dynamics (MD) simulations. Both computational studies agree with experiments near room temperature but deviate at elevated temperatures due to the neglect of anharmonic effects \cite{baroni2001phonons}.

\subsubsection{La--S system}

For the La--S system, Bolgar et al.~\cite{bolgar1987enthalpy} measured heat capacity, entropy, and enthalpy increment of lanthanum sulfide compounds covering a range of \ch{La:S} ratios using differential calorimetry. Vasilev et al.~\cite{vasilev1983physical} examined the low-temperature physical properties of rare-earth monosulfides, including \ch{LaS}, using a vacuum adiabatic calorimeter over a temperature range from \SI{2}{} to \SI{300}{K}. Their results for heat capacity are in good agreement with those of Bolgar et al.~\cite{bolgar1987enthalpy} at room temperature. The formation enthalpy of \ch{LaS} have been investigated by Semenkovich et al.~\cite{semenkovich1972enthalpies} using solution reaction calorimetry.

For compound $\gamma$-\ch{La2S3}, besides the measurement by Bolgar et al.~\cite{bolgar1987enthalpy},  Amano et al.~\cite{amano1986high} measured the heat capacity and enthalpy increment using a copper block drop calorimeter. Westrum et al. \cite{westrum1989thermophysical} further investigated the low-temperature entropy and heat capacity using adiabatic calorimetry. The results from these studies \cite{bolgar1987enthalpy, amano1986high, westrum1989thermophysical} are consistent within the overlapping temperature ranges. More recently, Vasilyeva and Nikolaev~\cite{vasilyeva2010la2s3} measured the heat capacity and formation enthalpy using a sensitive static tensimetric method at room temperature and their heat capacity deviated about \SI{1.6}{\percent} from previous results \cite{bolgar1987enthalpy, westrum1989thermophysical}. 

For \ch{La3S4}, Amano et al.~\cite{amano1986high} reported the heat capacity and enthalpy increment. Compared to Bolgar et al.~\cite{bolgar1987enthalpy}, their heat capacity values differ by \SI{4}{\percent}  at room temperature, whereas the  enthalpy increment is  consistent. Viennois et al.~\cite{viennois2013physical} obtained formation enthalpy and heat capacity at room temperature with the latter in agreement with values from Bolgar et al.~\cite{bolgar1987enthalpy} within \SI{0.3}{\percent}. 

For \ch{LaS2}, experimental data are scarce with the only reports of heat capacity provided by Vasilyeva and Nikolaev \cite{vasilyeva2010la2s3} and Bolgar et al.~\cite{bolgar1987enthalpy}, who additionally measured enthalpy increment for the phase.

\subsubsection{Ga--S system}

An early estimation of the heat capacity of \ch{Ga2S3} was provided by Moiseev and Sesták~\cite{moiseev1995thermochemical} using the Neumann--Kopp rule. More recently, Růžička et al.~\cite{ruuvzivcka2024heat} performed a comprehensive experimental and computational investigation on heat capacity and entropy of \ch{Ga2S3}. The authors used two techniques for low- and high-temperature measurements: relaxation calorimetry from \SI{2}{} to \SI{302}{K} and Tian--Calvet calorimetry in the range \SI{271}{} to \SI{1252}{K}. They supported these measurements with DFT calculations of the heat capacity that showed good agreement with the experiment. Růžička et al.~\cite{ruuvzivcka2024heat} also reported entropy in the temperature range from \SI{0}{} to \SI{350}{K} and room temperature formation enthalpy based on DFT. Barin \& Platzki ~\cite{barin1989thermochemical} reported the formation enthalpy of \ch{Ga2S3} at room temperature using solution calorimetric measurements.

For \ch{GaS}, Sedmidubský et al.~\cite{sedmidubsky2019chemical} reported the heat capacity over a wide range of temperatures using calorimetric measurements supplemented by theoretical modeling of both heat capacity and formation enthalpy using the Debye--Einstein approach. Two devices over different temperature range were employed: physical property measurement dystem (PPMS) Evercool--II equipment from \SI{2}{} to \SI{300}{K} and Micro DSC--III calorimeter from \SI{258}{} to \SI{358}{K}. They also provided heat capacity and formation enthalpy with theoretical modeling based on the Debye--Einstein approach. 
Hahn et al.~\cite{hahn1956uber} also reported the room temperature formation enthalpy of \ch{GaS}.

\section{Methods}

\subsection{Ab initio molecular dynamics}

To obtain thermodynamic properties of the liquid phases with scare experimental data, we performed AIMD simulations within the framework of density functional theory (DFT) using the Vienna Ab initio Simulation Package (VASP) \cite{kresse1993ab}. The electronic wave functions were expanded in a plane-wave basis set under periodic boundary conditions \cite{kresse1996efficiency}, with a plane-wave cutoff energy of \SI{400}{eV}. The exchange--correlation functional was described by the generalized gradient approximation \cite{perdew1992atoms}. The projector augmented wave method was used to describe the interaction between valence electrons and ion cores \cite{kresse1999ultrasoft}. Brillouin zone sampling was restricted to the $\Gamma$-point. All simulations were carried out in the canonical NVT ensemble with the temperature controlled by a Nosé--Hoover thermostat \cite{hoover1985canonical, nose1984unified}. Newton’s equations of motion were integrated using the Verlet algorithm with a time step of \SI{2}{fs}. The final 1000 configurations were used to statistically evaluate the thermodynamic and structural properties. 

Thermodynamic properties were calculated for the Ba--La, La--\ch{LaS}, and Ga--\ch{GaS} liquids. Ba–S and the sulfur-rich regions of the La--S and Ga--S systems were not considered because, at the temperatures of interest, the pure components are volatile: \ch{BaS} melts near \SI{2500}{K} \cite{boury2021liquid}, well above the boiling point of Ba (\SI{2118}{K} at 1 atm) \cite{greenwood2012chemistry}, while sulfur boils at \SI{718}{K} \cite{greenwood2012chemistry}. 
The simulations analyzed Ba--La and Ga--\ch{GaS} at \SI{1600}{K} and La--\ch{LaS} at \SI{3200}{K}. These temperatures were selected above the melting points and below the boiling points of relevant phases to ensure the evaluation of their liquid state. All supercells were constructed as cubic cells containing 128 atoms. The mixture compositions were specified at atomic fraction $\{ 0.2, 0.4, 0.5, 0.6, 0.8 \}$, along with pure reference states. The supercell volumes were fine-tuned to maintain the external pressure within $\pm$\SI{5}{kbar}, thereby minimizing the influence of pressure on the structure and total energy of the systems. To confirm that the chosen temperature was sufficient for melting and that the simulation time was long enough to reach equilibrium, pair correlation functions (PCFs) were calculated at selected stable time steps. The mixing enthalpy $\Delta H_\text{mix}$ was calculated as follows (taking the Ba--La system as an example):

\begin{equation}
\Delta H_\text{mix} = 
\frac{ E_\text{mix} - N_\text{Ba} \cdot E_\text{Ba} - N_\text{Ga} \cdot E_\text{Ga} }
     { N_\text{Ba} + N_\text{Ga} },
\end{equation}

\noindent where $E_\text{mix}$ is the average energy of the mixture supercell, $N_{\text{Ba}}$ and $N_{\text{Ga}}$ are the numbers of Ba and Ga atoms in the supercell, and $E_{\text{Ba}}$ and $E_{\text{Ga}}$ are the energies of pure Ba and pure Ga per atom, respectively.

\subsection{Thermodynamic modeling}
\label{sec:models}

For the pure elements (Ga, Ba, La, and S), we adopted the Gibbs free energy functions from the SGTE database \cite{dinsdale1991sgte}. For the binary and ternary phases, we selected appropriate models that describe substitutional solutions, stoichiometric compounds, liquid phases with short-range order, and solid phases with vacancies and substitutions. These models and their calibration to the experimental data  (reviewed in \Cref{sec:lit}) are described below. 

\subsubsection{Substitutional solutions}

For all liquid phases as well as solid b.c.c.-Ba, f.c.c.-La, and b.c.c.-La phases, we adopted a substitutional solution model suitable for random (liquid and solid) solutions. This model describes the molar Gibbs free energy, $G^\phi$, of a solution phase $\phi$ using the Redlich--Kister--Muggianu polynomial \cite{redlich1948thermodynamics}. For the b.c.c. system as an example:

\begin{equation}\label{eqn:rkm}
\begin{aligned}
    G^{b.c.c.}  &= \displaystyle\sum_{i=\text{Ba,La}} x_i {}^\circ G^\text{b.c.c.}_i \\
    & + RT \displaystyle\sum_{i=\text{Ba,La}}x_i\ln(x_i) \\
    & + x_\text{Ba}x_\text{La}\sum_{\theta=0}{}^\theta{}L^\text{b.c.c.}_\text{Ba,La}(x_\text{Ba}-x_\text{La})^\theta, 
\end{aligned}
\end{equation}

\noindent where $x_i$ denotes the molar fraction, ${}^\circ G^\text{b.c.c.}_i$ denotes the Gibbs free energy in the reference state of the b.c.c.\ phase, $R$ is the gas constant, $T$ is the temperature in Kelvin. The interaction parameters, ${}^\theta L^\text{b.c.c.}_\text{Ba,La}$ ($\theta \in [0,4]$), are considered to linearly depend on temperature: ${}^\theta L^\text{b.c.c.}_\text{Ba,La} = {}^\theta A^\text{b.c.c.}_\text{Ba,La} + {}^\theta B^\text{b.c.c.}_\text{Ba,La} \cdot T$. The parameters ${}^\theta A^\text{b.c.c.}_\text{Ba,La}$ and ${}^\theta B^\text{b.c.c.}_\text{Ba,La}$ shall be calibrated to the experimental data.

\subsubsection{Stoichiometric compounds}

The Ba--S, La--S, and Ga--S binary systems include stable stoichiometric compounds (such as \ch{BaS}, \ch{BaS2}, \ch{BaS3}, \ch{LaS}, \ch{GaS}, \ch{Ga2S3}). For the compounds that lack published thermodynamic information, we estimate the molar Gibbs free energies using the Neumann--Kopp rule \cite{kopp1865iii, gong2024revisiting, zhou2014first}:

\begin{equation}
\begin{split}
G_f^{M_xN_y} &= x \cdot ^oG_M + y \cdot ^oG_N + A_{M_xN_y} \\
&+ B_{M_xN_y} \cdot T, 
\end{split}
\end{equation}

\noindent where $x$ and $y$ are the stoichiometric coefficients and $A_{M_xN_y}$ and $B_{M_xN_y}$ are the adjustable parameters for compounds $M_xN_y$. For the compounds with available experimental heat capacity data, higher order parameters are included to describe the deviation from Neumann--Kopp rule of the measured heat capacity:

\begin{equation}
\begin{aligned}
    G_f^{M_xN_y} &= x \cdot ^oG_M + y \cdot ^oG_N + A_{M_xN_y} \\ 
    &+ B_{M_xN_y} \cdot T + C_{M_xN_y} \cdot T \cdot \ln(T) \\
    &+ D_{M_xN_y} \cdot T^{-1} + E_{M_xN_y} \cdot T^2. 
\end{aligned}
\end{equation}

\subsubsection{Liquid phases with short-range order} 

In the Ba--S, La--S, and Ga--S binary phase diagrams, the melting points of compounds are much higher than those of the corresponding pure elements. We therefore assume that, in the liquid phase, the compounds retain strong attractive interactions even at temperatures higher than their melting points. Consequently, short-range chemical order is likely to occur in the liquid state \cite{schmid1985thermodynamic}. The associated solution model was chosen for the \ch{BaS}, \ch{LaS}, \ch{La2S3}, \ch{GaS}, and \ch{Ga2S3} phases. As a specific example, the Gibbs energy of \ch{La2S3} in the liquid state is written as: 

\begin{equation}
G^\text{Liq} = \sum_{i} x_i G^\text{Liq}_i + RT \sum_{i} x_i\ln(x_i) + G^\text{Liq}_\text{exc},
\end{equation}

\noindent where $x_i$ is the mole fraction of the $i^{th}$ species ($i \in$ \{\ch{La}, \ch{S}, \ch{La2S3}\}) in the liquid phase and  $G^\text{Liq}_\text{exc}$ is the excess Gibbs free energy expressed as:

\begin{equation}
G^\text{Liq}_\text{exc} = \sum_i \sum_{j \neq i} x_ix_j \sum_{\theta=0}^n {}^\theta L_{i,j}^\text{Liq}(x_i-x_j)^\theta
\end{equation}

\noindent with $i,j \in \{\ch{La}, \ch{S}, \ch{La2S3}\}$. The binary interaction parameters are linear functions of temperature ${}^\theta L_{i,j}^\text{Liq} = {}^\theta A_{i,j}^\text{Liq} + {}^\theta B_{i,j}^\text{Liq} \cdot T$, the parameters ${}^\theta A_{i,j}^\text{Liq}$ and ${}^\theta B_{i,j}^\text{Liq}$ are to be calibrated.

\subsubsection{Solid solution phase with vacancies and substitution} 

Based on literature sources \cite{boury2021liquid, zhang2017structural, kamarzin1981growth}, the non-stoichiometry of the $\gamma$-\ch{La2S3} phase arises from the presence of vacancies on the La sublattice. The actual composition, accounting for vacancies ($Va$), is expressed as $\gamma$-\ch{La_{2.25-x}}$Va$\ch{_xS3}, where $0 \leq \ch{x} \leq 0.25$. This range corresponds to the end-members $\gamma$-\ch{La_{2.25}S3} (\ch{La3S4}) and $\gamma$-\ch{La2S3}. To describe the $\gamma$-\ch{La2S3} phase considering vacancies, the sublattice model can be \ch{(La)2(La,$Va$)_{0.25}(S)3}. For the Ba--La--S ternary system, literature data \cite{andreev1991interaction} report approximately \SI{50}{\percent} solid solubility of \ch{BaS} in $\gamma$-\ch{La2S3}, resulting in the formation of an intermediate phase with an approximate composition of \ch{Ba_{0.75}La_{1.5}S3} (\ch{BaLa2S4}). Although the detailed substitution mechanism is not specified, it is assumed that partial substitution of La by Ba accounts for the observed composition. Accordingly, the sublattice model used in the Ba--La--S database for this phase is \ch{(La)_{1.25}(La,Ba)_{0.75}(La,$Va$)_{0.25}(S)3}. In contrast, for the Ga--La--S ternary system, no significant solubility of $\gamma$-\ch{La2S3} in the \ch{Ga2S3}--\ch{La2S3} pseudo-binary system has been reported. Therefore, the original sublattice model \ch{(La)2(La,$Va$)_{0.25}(S)3} is retained for $\gamma$-\ch{La2S3} in this system.
For Ba--La--S system, the three terms of the $\gamma$-\ch{La2S3} Gibbs energy functions are:

\begin{subequations}
\begin{equation}
\begin{aligned}
    G^{\gamma-\ch{La2S3}}_\text{ref} &= y_{\text{La}}^{\prime}y_{Va} ^{\prime\prime} G^{\gamma-\ch{La2S3}}_{\text{La:La:$Va$:S}} \\
    &+ y_{\text{La}}^{\prime} y_{\text{La}}^{\prime\prime} G^{\gamma-\ch{La2S3}}_{\text{La:La:La:S}} \\
    & + y_{\text{Ba}}^{\prime}y_{Va} ^{\prime\prime} G^{\gamma-\ch{La2S3}}_{\text{La:Ba:$Va$:S}} \\
    &+ y_{\text{Ba}}^{\prime}y_{\text{La}} ^{\prime\prime} G^{\gamma-\ch{La2S3}}_{\text{La:Ba:La:S}},
\end{aligned}
\end{equation}

\begin{equation}
\begin{aligned}
    G^{\gamma-\ch{La2S_3}}_\text{ide} = \\
    RT[0.75\left(y_{\ch{\text{La}}}^{\prime}\ln(y_{\text{La}}^{\prime}) + y_{\text{Ba}}^{\prime}\ln(y_{\text{Ba}}^{\prime})\right) \\ 
     + 0.25\left(y_{\ch{\text{La}}}^{\prime\prime}\ln(y_{\text{La}}^{\prime\prime}) + y_{Va}^{\prime\prime}\ln(y_{Va}^{\prime\prime})\right)],
\end{aligned}
\end{equation}

\begin{equation}
\begin{aligned}
    G^{\gamma-\ch{La2S3}}_\text{exc} = \\
    y_{\text{La}}^{\prime}y_{\text{Ba}}^{\prime}y_{\text{La}}^{\prime\prime} \sum_{\theta=0}^n{}^\theta L_{\text{La:La,Ba:La:S}}^{\gamma-\ch{La2S3}}(y_{La}^{\prime}-y_{\text{Ba}}^{\prime})^\theta \\
    + y_{\text{La}}^{\prime}y_{\text{Ba}}^{\prime}y_{Va}^{\prime\prime} \sum_{\theta=0}^n{}^\theta L_{\text{La:La,Ba:$Va$:S}}^{\gamma-\ch{La2S3}}(y_{\text{La}}^{\prime}-y_{\text{Ba}}^{\prime})^\theta \\
    + y_{\text{La}}^{\prime}y_{\text{La}}^{\prime\prime}y_{Va}^{\prime\prime} \sum_{\theta=0}^n{}^\theta L_{\text{La:La:La:$Va$:S}}^{\gamma-\ch{La2S3}}(y_{\text{La}}^{\prime\prime}-y_{Va}^{\prime\prime})^\theta \\
    + y_{\text{Ba}}^{\prime}y_{\text{La}}^{\prime\prime}y_{Va}^{\prime\prime} \sum_{\theta=0}^n{}^\theta L_{\text{La:Ba:La:$Va$:S}}^{\gamma-\ch{La2S3}}(y_{\text{La}}^{\prime\prime}-y_{Va}^{\prime\prime})^\theta, 
\end{aligned}
\end{equation}
\end{subequations}

\noindent where $y_{\text{La}}^{\prime}$ and $y_{\text{Ba}}^{\prime}$ are the site fractions of La and Ba atoms in the first sublattice, $y_{\text{La}}^{\prime\prime}$ and $y_{Va}^{\prime\prime}$ are the site fraction of La atoms and vacancies in the second sublattice. $G^{\gamma-\ch{La2S3}}_{\text{La:La:$Va$:S}}$, $G^{\gamma-\ch{La2S3}}_{\text{La:La:La:S}}$ and $G^{\gamma-\ch{La2S3}}_{\text{La:Ba:$Va$:S}}$ are the Gibbs energy of the compounds \ch{La2S3}, \ch{La3S4} (\ch{La_{2.25}S3}), \ch{BaLa2S4} (\ch{La_{1.5}Ba_{0.75}S3}). $G^{\gamma-\ch{La2S3}}_{\text{La:Ba:La:S}}$ is the Gibbs energy of the hypothetical compound \ch{La_{1.25}Ba_{0.75}S3}.

\begin{figure*}[h]
  \centering
  \includegraphics[width=0.75\linewidth]{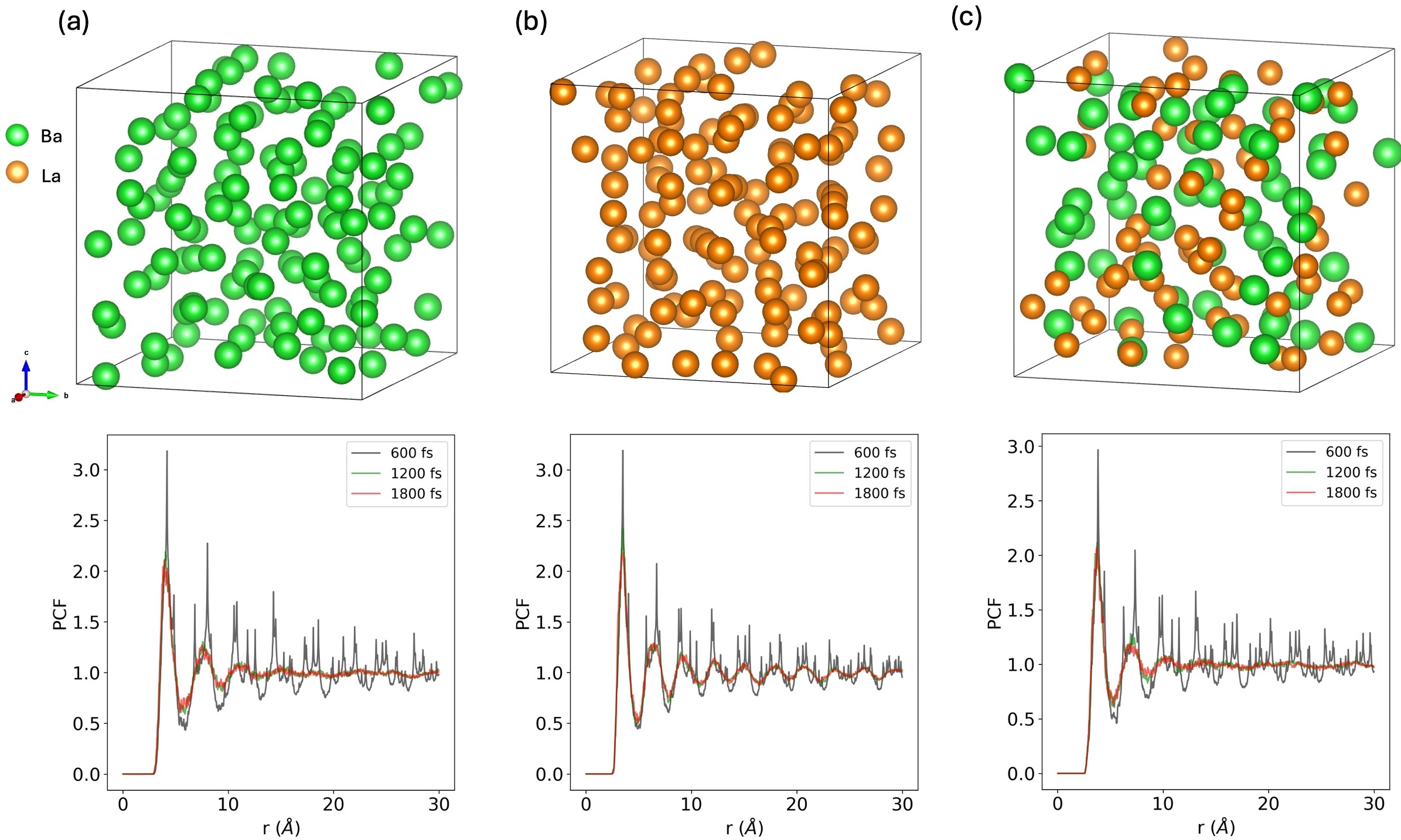}
  \caption{Representative equilibrium snapshots of the supercells for (a) pure Ba liquid, (b) pure La liquid, and (c) the 50~at.\SI{}{\percent} Ba--La liquid alloy at \SI{1600}{K}, as well as their corresponding PCFs at three time steps.}
  \label{fig:BaLa_supercell}
\end{figure*}

\subsection{Optimization methodology}

We used the described models to calibrate thermodynamic descriptions for all the phases identified in the literature for the Ba--La--S and Ga--La--S systems (\Cref{tab:Ba-La-S_phases,tab:Ga-La-S_phases}). Specifically, we fitted the parameters in the Gibbs free energy functions to the experimental data reviewed in \Cref{sec:lit} as well as AIMD calculations carried out in this study for the liquid phases and described in \Cref{sec:aimd_results}. The parameters were optimized in the PanOptimizer module of the Pandat software \cite{chen2002pandat}.

\section{Results and discussion}

\subsection{AIMD calculations of liquid phases}
\label{sec:aimd_results}

\begin{figure*}[h]
  \centering
  \includegraphics[width=\linewidth]{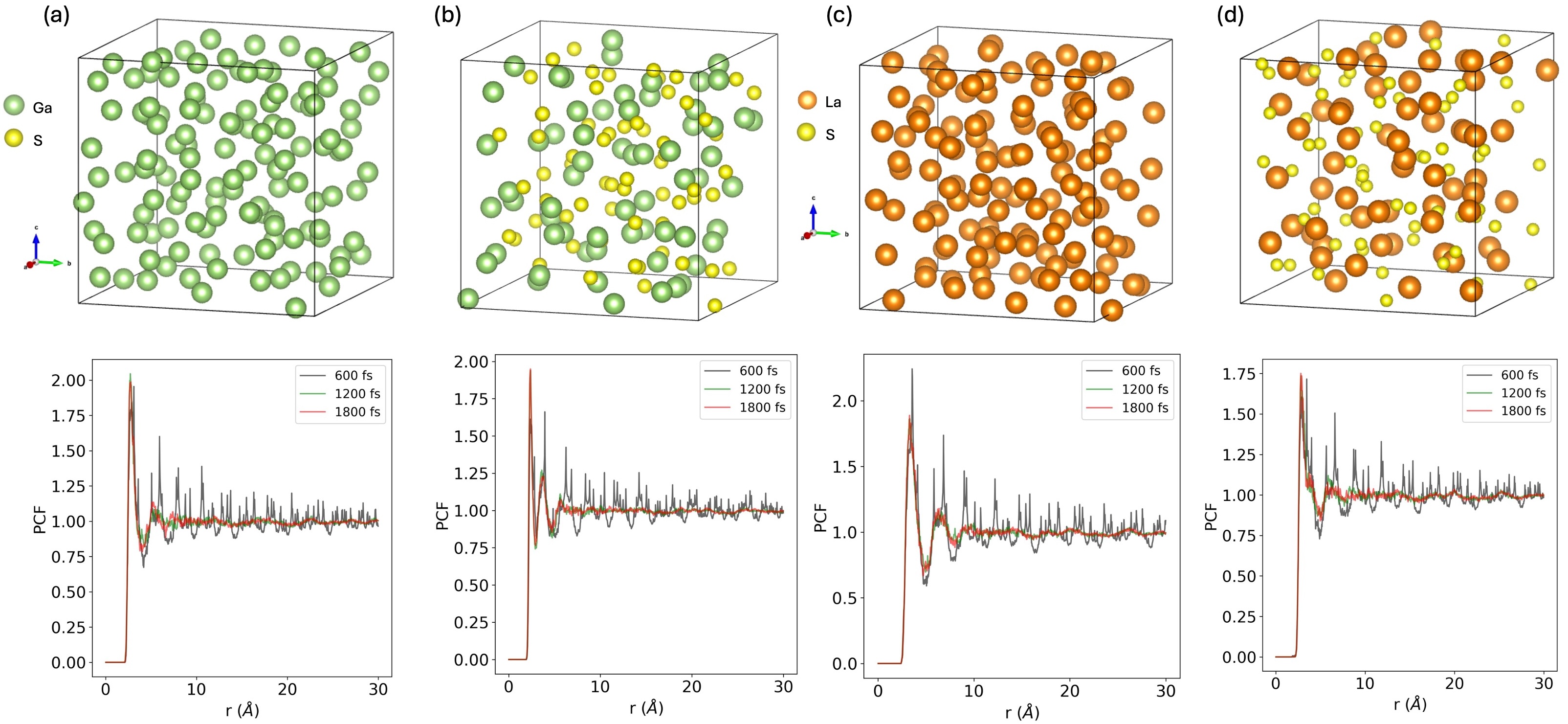}
  \caption{Representative equilibrium snapshots of the supercells for (a) pure Ga and (b) \ch{GaS} liquids at \SI{1600}{K}; (c) pure La and (d) \ch{LaS} liquids at \SI{3200}{K} as well as their corresponding PCFs at three time steps.}
  \label{fig:GaS_LaS_supercell}
\end{figure*}

Snapshots of equilibrated structures and the corresponding PCFs (\Cref{fig:BaLa_supercell,fig:GaS_LaS_supercell} confirm the liquid state has been reached as targeted in the AIMD simulations. Peaks corresponding to the long-range order broaden with time and disappear by simulation time step of \SI{1200}{\femto\second}. The coincidence of the PCF curves at \SI{1800}{\femto\second} and \SI{1200}{\femto\second} further confirms the equilibrium condition reached within the time considered in the AIMD simulations. 

\begin{figure*}[h]
  \centering
  \includegraphics[width=\linewidth]{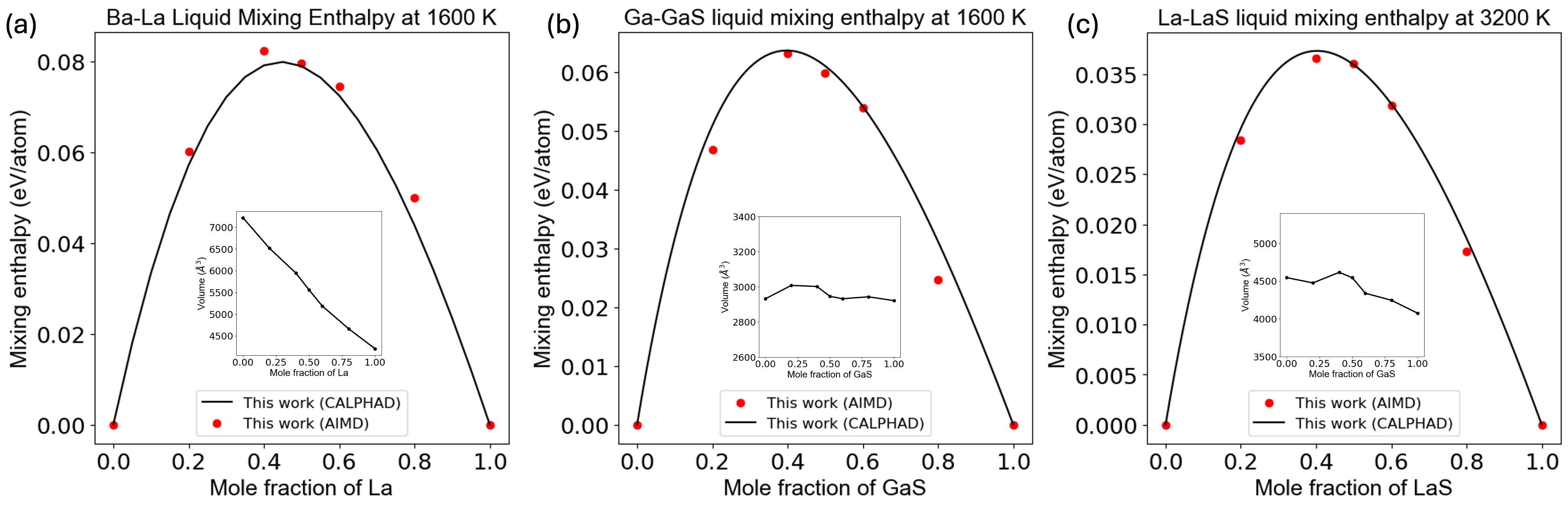}
  \caption{Mixing enthalpy of alloy liquids at various compositions for (a) Ba--La at \SI{1600}{K}, (b) Ga--\ch{GaS} at \SI{1600}{K}, and (c) La--\ch{LaS} at \SI{3200}{K}. The mixing enthalpy from the present AIMD simulations (red symbols) and the ensuing CALPHAD assessment (black curves) are seen to be consistent. The inset shows the equilibrium supercell volumes at different compositions under nearly zero internal pressure.}
  \label{fig:mixing_enthalpy}
\end{figure*}

Upon confirmation that equilibrium was reached in the liquid phases using AIMD, we calculated the mixing enthalpy of Ba--La, Ga--\ch{GaS}, and La--\ch{LaS} systems obtained with the simulations (symbols in \Cref{fig:mixing_enthalpy}). All three systems show that mixing is energetically unfavorable compared to the pure components at the studied temperatures. The asymmetric composition dependence of the mixing enthalpy reflects the non-ideal interactions in the liquids. These mixing enthalpy values are adopted for fitting the interaction parameters of the relevant phases in our assessment and evaluating the assessment results (see \Cref{sec:assessment}). Additionally, the equilibrium supercell volumes at nearly zero internal pressure from AIMD simulations are presented (see inset in \Cref{fig:mixing_enthalpy}). 

\subsection{Thermodynamic assessment}
\label{sec:assessment}

All phase diagrams obtained with the CALPHAD method in this study are presented in \Cref{fig:Ba-La-S_phase_diagram} for the Ba--La--S system and \Cref{fig:Ga-La-S_phase_diagram} for the Ga--La--S system along with the literature data available for these phase diagrams. \Cref{tab:Ba-La-S_invariant_reaction,tab:Ga-La-S_invariant_reaction} lists the invariant reactions in the two systems -- both calculated in this study and reported in the literature. The sets of functions with calibrated parameters underlying these results are given in \Cref{tab:Ba-La-S_solid_TDB,tab:Ba-La-S_liq_TDB,tab:Ga-La-S_solid_TDB,tab:Ga-La-S_liq_TDB}. The thermodynamic properties calculated with these calibrated models are shown in \Cref{fig:Ba-S_thermo_prop,fig:Ga-S_thermo_prop,fig:La-S_thermo_prop}.  Overall, we observe good agreement between our calculations and data reported in the literature. Below, we discuss these results in more detail for both systems. 

\subsubsection{The Ba--La--S system}

\begin{figure*}[h]
  \centering
  \includegraphics[width=0.6\linewidth]{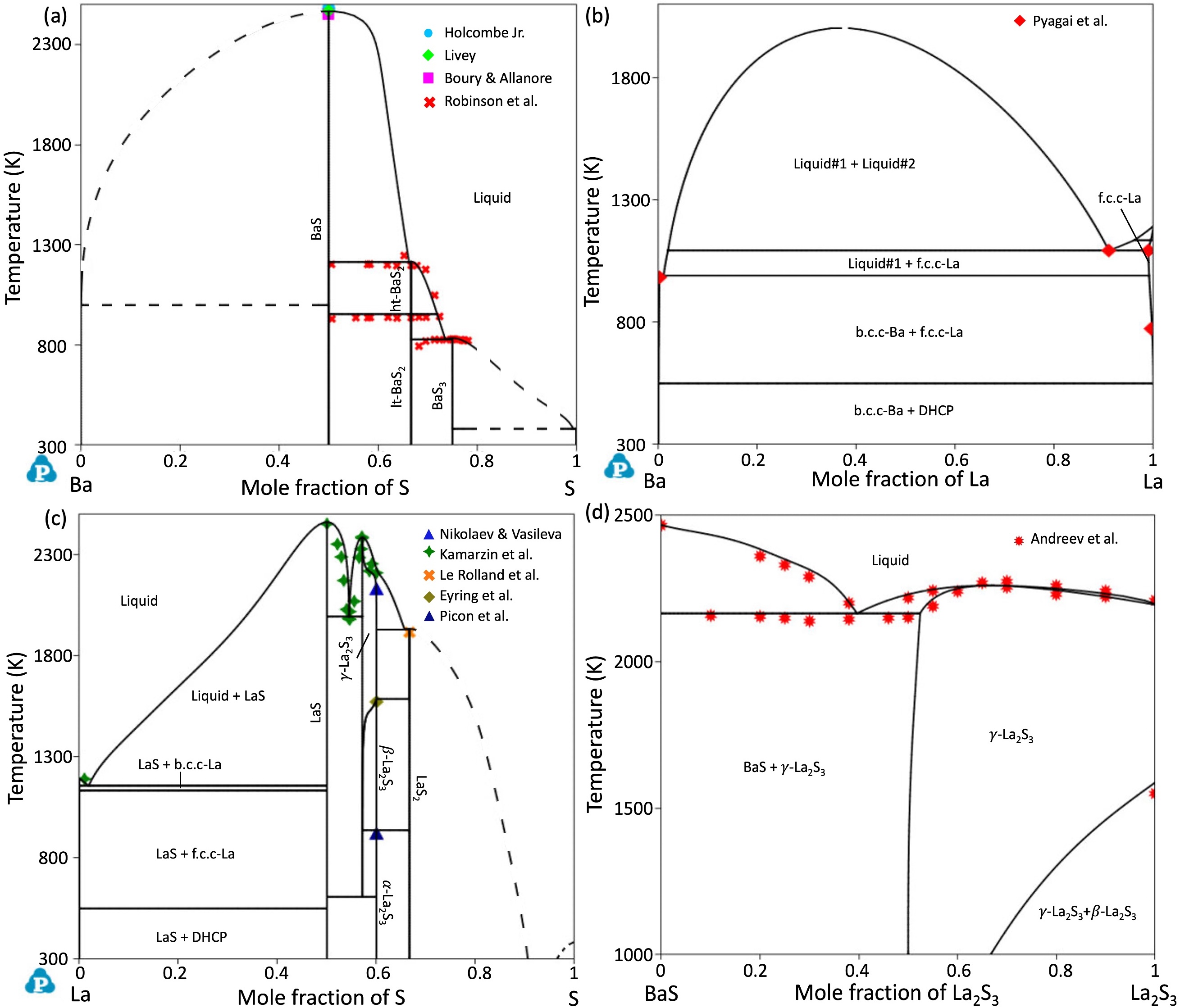}
  \caption{Phase diagrams calculated for the Ba--La--S system along with the literature data: (a) Ba--S; (b) Ba--La; (c) La--S; (d) \ch{BaS}--\ch{La2S3}. Symbols overlaid with the phase diagram lines represent literature data. The dashed lines represent calculated liquidus lines in regions currently lacking experimental data (and thus require further validation).}
  \label{fig:Ba-La-S_phase_diagram}
\end{figure*}

In the composition range where experimental data are available ($0.5 < x_\text{S} < 0.8$) \cite{boury2021liquid, livey1959high, holcombe1984tentative, robinson1931xcv}, the calculated Ba--S binary phase diagram (\Cref{fig:Ba-La-S_phase_diagram}(a)) reproduces the key features: the congruent melting point of \ch{BaS}, the high-temperature phases ht-\ch{BaS2} and \ch{BaS3}, and the phase transition between ht- and lt-\ch{BaS2}. Since multiple melting temperatures have been reported for \ch{BaS}, we adopted the mean of the two consistent values from literature \cite{livey1959high, holcombe1984tentative}, i.e., \SI{2475}{K}. For rest of the Ba--S compounds, the assessment is calibrated to the only widely cited source \cite{robinson1931xcv}. To account for the experimentally observed steep liquidus line, strong associate interactions were introduced, enabling accurate reproduction of the available liquidus data and the corresponding solidus line.

The calculated Ba--La binary phase diagram (\Cref{fig:Ba-La-S_phase_diagram}(b)) reproduces the reported invariant reactions, including the limited solid solubilities in b.c.c.-Ba (0.2~at.\SI{}{\percent}), b.c.c.-La (1.0~at.\SI{}{\percent}), and f.c.c.-La, as well as the monotectic points at 0.4 and 90~at.\SI{}{\percent}. The mixing enthalpy at \SI{1600}{K} of the Ba--La liquid phase obtained in AIMD simulations  (\Cref{fig:mixing_enthalpy}(a)) was used as a supplement to the scarce experimental data. With the incorporation of AIMD constraints, the resulting CALPHAD description yields a phase diagram consistent with the only published experimental study  \cite{pegaj1985phase}.

For the La--S binary phase diagram (\Cref{fig:Ba-La-S_phase_diagram}(c)), experimental data are available up to 70~at.\SI{}{\percent} sulfur content. Our assessment relied mainly on the comprehensive measurements of Kamarzin et al.~\cite{kamarzin1981growth}, including the eutectic points between S and \ch{LaS} and between \ch{LaS} and $\gamma$-\ch{La2S3}, the eutectic liquidus line of \ch{LaS}--$\gamma$-\ch{La2S3}, the liquidus of the $\gamma$-\ch{La2S3} solid solution, and the congruent melting of \ch{LaS}. Since Kamarzin et al.\ did not report the melting of $\gamma$-\ch{La2S3}, we adopted the values from Nikolaev and Vasileva \cite{nikolaev2008vapor}. The $\alpha \leftrightarrow \beta$ and $\beta \leftrightarrow \gamma$ phase transitions of \ch{La2S3} were taken from Picon et al.~\cite{picon1956sous} and Eyring et al.~\cite{eyring2002handbook}, respectively. For \ch{LaS2}, the melting point reported by Le Rolland et al.~\cite{rolland1991vibrational} was used. Because the transition temperature between orthorhombic \cite{rolland1991vibrational} and monoclinic \cite{dugue1978structure} \ch{LaS2} is ambiguous in the literature, these polymorphs were treated as a single phase in our assessment. In addition, \ch{La3S4} is unstable at room temperature according to our results, which is consistent with experimental observations \cite{picon1956sous, schleid1991m10s14o} and with DFT calculations from the Materials Project \cite{jain2013commentary}. 

Both Ba--S and La--S phase diagrams include liquidus lines obtained according to the present assessment, which however currently lack experimental data. These liquidus lines may need further experimental validation (dashed lines in \Cref{fig:Ba-La-S_phase_diagram}(a,c)).

For the ternary Ba--La--S system, we focus on the \ch{BaS}--\ch{La2S3} pseudo-binary phase diagram because it is the only section that has been experimentally investigated in the literature. For our assessment, we followed the comprehensive study by Andreev et al.~\cite{andreev1991interaction}, who reported an incongruent melting point for $\gamma$-\ch{La2S3} and large solubility of \ch{BaS} in $\gamma$-\ch{La2S3}. The melting points of both \ch{BaS} and $\gamma$-\ch{La2S3} used in this work are consistent with other references~\cite{kamarzin1981growth, massalski1990binary}. A pseudo-binary phase diagram calculated according to these data is shown in \Cref{fig:Ba-La-S_phase_diagram}(d).

We next present the thermodynamic properties of solid phases pertinent to the Ba--La--S system obtained as part of our assessment and their comparison with available experimental measurements and AIMD calculations.

\begin{figure}[H]
  \centering
  \includegraphics[width=\linewidth]{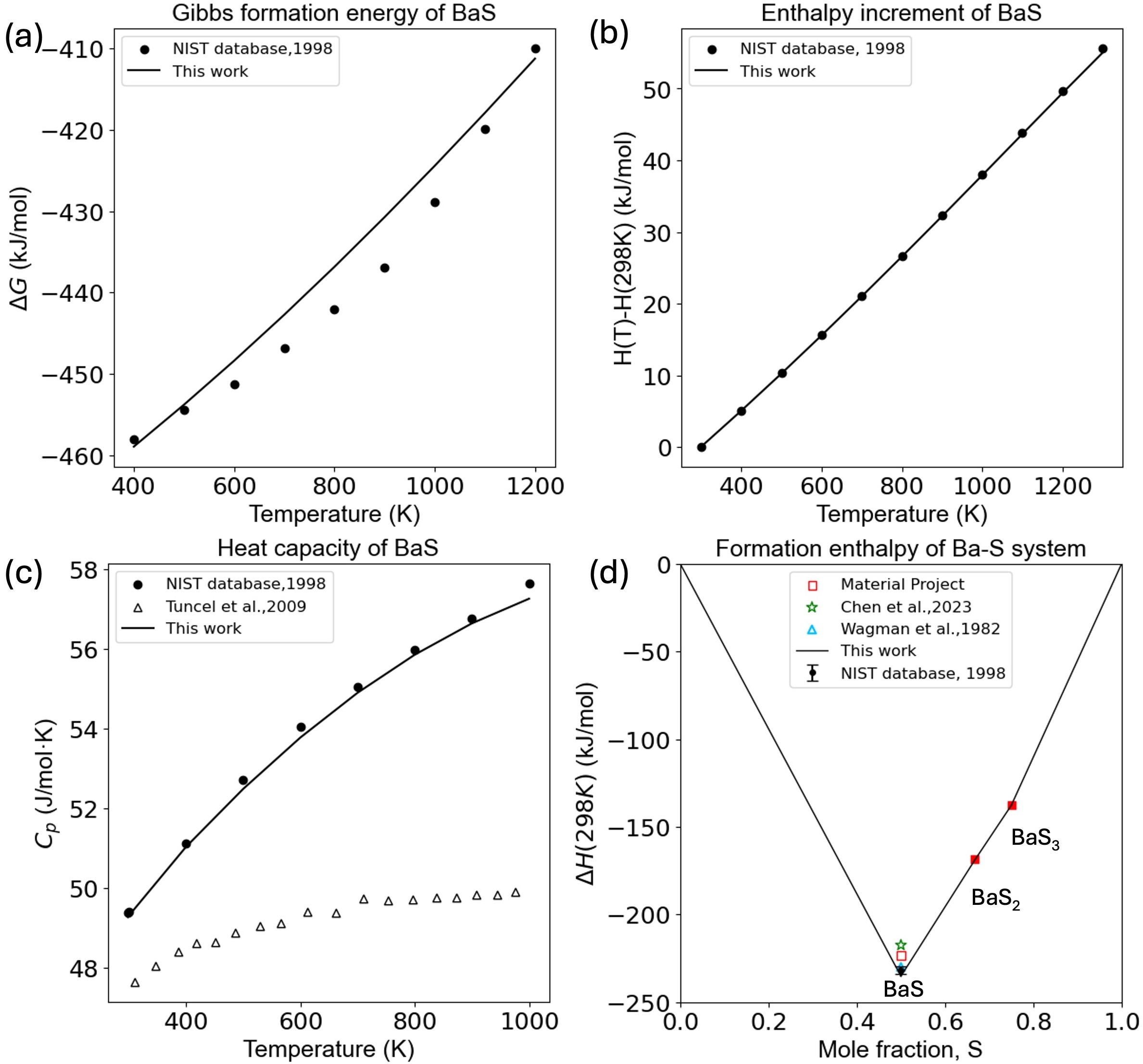}
  \caption{Thermodynamic properties for the Ba--S binary system: (a) formation Gibbs free energy, (b) enthalpy increment, and (c) heat capacity of \ch{BaS}; (d) formation enthalpy of the Ba--S phases. Symbols represent literature data: filled for data included in the parameter calibration, open symbols for data shown only for comparison.}
  \label{fig:Ba-S_thermo_prop}
\end{figure}

\Cref{fig:Ba-S_thermo_prop} presents the thermodynamic properties of the Ba--S system. The Gibbs formation energy and enthalpy increment of \ch{BaS} over the temperature range \SIrange{400}{1200}{K} obtained in our assessment show a good fit (\Cref{fig:Ba-S_thermo_prop}(a,b)) to the data from the NIST database~\cite{chase1998nist}, which served as the only source of experimental measurements of these properties for our model calibration. The heat capacity of \ch{BaS} (\Cref{fig:Ba-S_thermo_prop}(c)) is compared to the experimental values from the NIST database~\cite{chase1998nist} and the first-principles calculations by Tuncel et al.~\cite{tuncel2009first} over the temperature range \SIrange{350}{1000}{K}. Since DFT predictions neglected anharmonic contributions (as discussed in \Cref{sec:lit}) and deviated from experiments at elevated temperatures, we only used the NIST data for parameter calibration and show DFT results for comparison.

\Cref{fig:Ba-S_thermo_prop}(d) shows the formation enthalpies at \SI{298}{K} for Ba--S compounds, including \ch{BaS}, low-temperature \ch{BaS2}, and \ch{BaS3}. For \ch{BaS}, the NIST database  reports $\Delta H(298K) = -231.8\pm2.1$ \SI{}{kJ\per\mol} (shown as error bar in \Cref{fig:Ba-S_thermo_prop}(d)), which is consistent with our assessment giving  the value of $-232.2$ \SI{}{kJ\per\mol}. 
Additional reference data from the Materials Project~\cite{jain2013commentary}, NBS~\cite{wagman1982nbs}, and recent DFT studies~\cite{chen2023theoretical} are included for comparison. 

\begin{figure}[h]
  \centering
  \includegraphics[width=\linewidth]{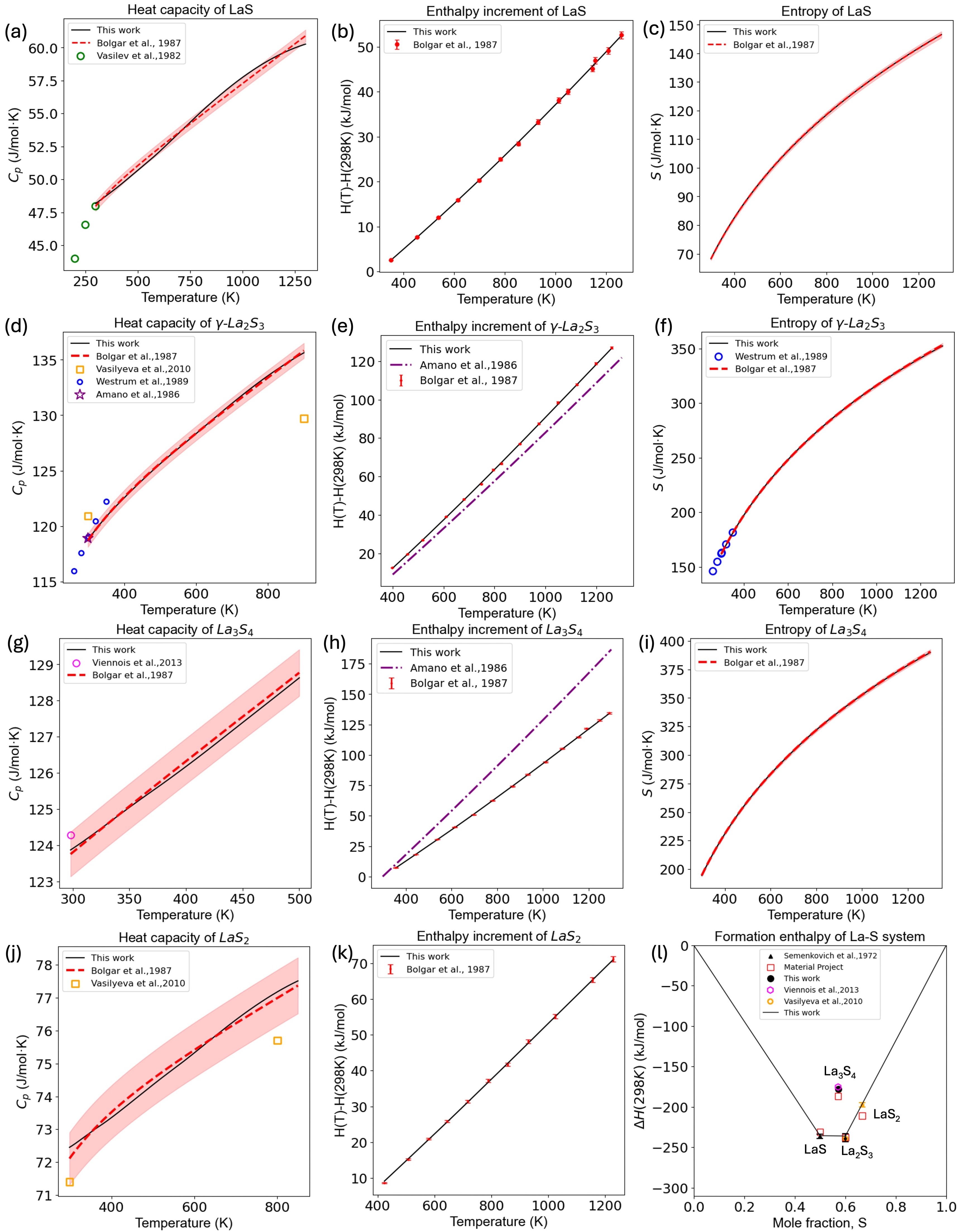}
  \caption{Thermodynamic properties for the La--S binary system: heat capacity, enthalpy increment, and entropy of (a--c) \ch{LaS}; (d--f) $\gamma$-\ch{La2S3}; (g--i) \ch{La3S4}; (j--k) heat capacity and enthalpy increment of \ch{LaS2} and formation enthalpy for the La--S phases. The shaded regions represent measurement uncertainty. Symbols represent literature data: filled for data included in the parameter calibration, open symbols for data shown only for comparison. Functional dependencies from literature are represented by lines: ``dashed'' for functions included in calibration, ``dash-dotted'' shown for reference only.}
  \label{fig:La-S_thermo_prop}
\end{figure}

Thermodynamic properties of solid phases in the La--S binary system are summarized in \Cref{fig:La-S_thermo_prop}, including the heat capacity, enthalpy increment, entropy and formation enthalpy from literature and our database. The properties of La--S compounds reported by Bolgar et al.~\cite{bolgar1987enthalpy} were mainly used for the assessment, and our database match well with the experimental data. 
Enthalpy increments were measured by Bolgar et al. \cite{bolgar1987enthalpy} using high-temperature differential calorimeter with  the measurement uncertainties of \SI{0.75}{\percent} for \ch{LaS}, \SI{0.5}{\percent} for \ch{La2S3}, \SI{0.5}{\percent} for \ch{La3S4} and \SI{1.1}{\percent} for \ch{LaS2};  these uncertainties were 
propagated to the derived heat capacity and entropy, $C_{p}(T)$ and $S(T)$. 

\Cref{fig:La-S_thermo_prop}(a--c) shows the thermodynamic properties of \ch{LaS}, including low-temperature heat capacity from Vasilev et al.~\cite{vasilev1983physical} obtained with vacuum adiabatic calorimeter (\Cref{fig:La-S_thermo_prop}(a)). The two heat capacity sources \cite{bolgar1987enthalpy, vasilev1983physical} cover different temperature ranges but overlap at room temperature, where their values agree. The enthalpy increment and entropy of \ch{LaS} were fitted to the results of Bolgar et al. as the only available data source (\Cref{fig:La-S_thermo_prop}(b,c)).

\Cref{fig:La-S_thermo_prop}(d--f) show the properties of $\gamma$-\ch{La2S3}. Several experimental studies report the heat capacity \cite{bolgar1987enthalpy, vasilyeva2010la2s3, westrum1989thermophysical, amano1986high} with good consistency around room temperature except the measurements by Vasilyeva and Nikolaev~\cite{vasilyeva2010la2s3} at \SI{900}{K}. 
The high-temperature discrepancy likely reflects methodological differences: Vasilyeva and Nikolaev~\cite{vasilyeva2010la2s3} infer heat capacities indirectly from vapor-pressure equilibria that is sensitive to sulfur nonstoichiometry and volatility. Whereas Bolgar et al.~\cite{bolgar1987enthalpy} report direct high-temperature calorimetry on phase-verified \ch{La2S3} with stated uncertainties, which we adopt as the more reliable basis. 
\Cref{fig:La-S_thermo_prop}(e,f) show the enthalpy increment and entropy of $\gamma$-\ch{La2S3} which are consistent across studies for a wide range of temperatures. Our assessment therefore adopted representative data by Bolgar et al.~\cite{bolgar1987enthalpy} for parameter calibration, which results in a good agreement between the calculated and literature values (\Cref{fig:La-S_thermo_prop}(e,f)).

For \ch{La3S4}, we adopted the thermodynamic property data from Bolgar et al.\ \cite{bolgar1987enthalpy}, which are in agreement with our results upon parameter calibration (\Cref{fig:La-S_thermo_prop}(g--i)). The enthalpy increment measurements of Amano et al.\ \cite{amano1986high} align with those of Bolgar et al.\ at low temperatures but deviate significantly at higher temperatures. This discrepancy is likely due to impurities introduced by the preparation method employed by Amano et al., combined with their lack of detailed chemical composition data for the \ch{La3S4} samples. These reasons reinforce our choice for model calibration to the data from Bolgar et al. In addition, the DFT-calculated heat capacities of \ch{La3S4} by Viennois et al.\ \cite{viennois2013physical} show good agreement with Bolgar et al.\ at room temperature and are included in \Cref{fig:La-S_thermo_prop}(g).

For less investigated \ch{LaS2}, the heat capacity of \ch{LaS2} was reported by Vasilyeva and Nikolaev \cite{vasilyeva2010la2s3} and by Bolgar et al.\ \cite{bolgar1987enthalpy}.  As shown in \Cref{fig:La-S_thermo_prop}(j), the two datasets are consistent at low temperatures but diverge at higher temperatures. Vasilyeva and Nikolaev report their value for \SI{900}{K} by analogy to similar systems (\ch{PrS2}–\ch{SmS2}) rather than from direct measurements. For this reason and because Bolgar et al.\ cover a broader temperature range, we adopted the data of Bolgar et al.\ in our assessment. Furthermore, as shown in \Cref{fig:La-S_thermo_prop}(k), our database calibrated the enthalpy increments to the data of Bolgar et al.\ and exhibits excellent agreement.

\Cref{fig:La-S_thermo_prop}(l) compares our calculated formation enthalpies of La--S phases at \SI{298}{K} with literature data \cite{jain2013commentary, semenkovich1972enthalpies, viennois2013physical, vasilyeva2010la2s3}. The formation enthalpies of \ch{LaS} and $\gamma$-\ch{La2S3} were calibrated to the experimental measurements of Semenkovich et al.\ \cite{semenkovich1972enthalpies} with measurement uncertainties of $\pm\SI{2.5}{kJ.mol^{-1}}$ and $\pm\SI{5}{kJ.mol^{-1}}$, respectively. For \ch{LaS2}, the value was calibrated to Vasilyeva et al.\ \cite{vasilyeva2010la2s3} with $\pm\SI{2.6}{kJ.mol^{-1}}$ measurement uncertainty, as it represents the only available experimental dataset. In the case of \ch{La3S4}, the formation enthalpy agrees with the DFT results of Viennois et al.\ \cite{viennois2013physical} and lies above the convex hull, indicating its metastability at room temperature, which is consistent with the La--S phase diagram obtained in this study. For completeness, predictions from the Materials Project \cite{jain2013commentary} are also included in the comparison.

\subsubsection{The Ga--La--S system}

\begin{figure*}[h]
  \centering
  \includegraphics[width=\textwidth]{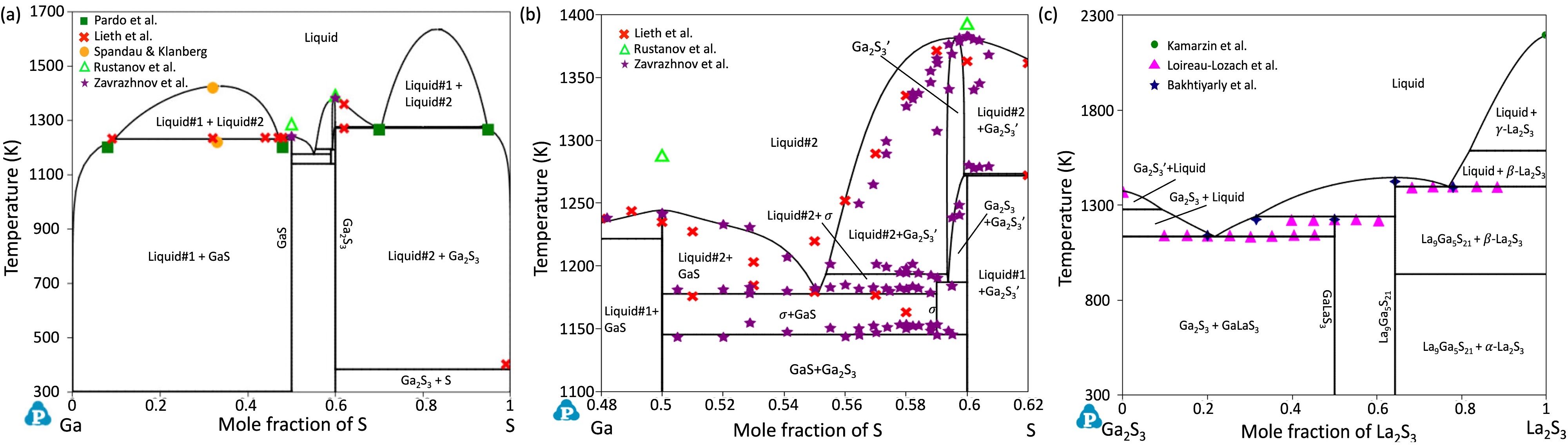}
  \caption{Phase diagrams calculated for the Ga--La--S system along with the literature data: (a) Ga--S and (b) its detailed view in the composition range (48 to 62~at.\SI{}{\percent} sulfur) above \SI{1100}{K}; (c) \ch{Ga2S3}--\ch{La2S3}. Symbols represent the literature data: filled for data used for the parameter calibration, open symbols for data shown only for comparison.}
  \label{fig:Ga-La-S_phase_diagram}
\end{figure*}

The thermodynamic parameters obtained in the present work for the Ga--La--S system are listed in \Cref{tab:Ga-La-S_solid_TDB,tab:Ga-La-S_liq_TDB}. One binary phase diagram (Ga--S) and one pseudo-binary phase diagram (\ch{Ga2S3}--\ch{La2S3}) were calculated using the developed thermodynamic database and are shown in \Cref{fig:Ga-La-S_phase_diagram}.

For the Ga--S binary phase diagram (\Cref{fig:Ga-La-S_phase_diagram}(a)), we adopted the experimental interpretation from widely cited studies \cite{pardo1993diagramme, lieth1966p, zlomanov1987ptx, greenberg2001thermodynamic, zavrazhnov2018phase, spandau1958thermische}, rather than the alternative version reported by Rustanov et al. \cite{rustanov1967investigation}. Pardo et al.~\cite{pardo1993diagramme} and Lieth et al.~\cite{lieth1966p} identified a miscibility gap on the Ga-rich side (from 10 to 48~at.\SI{}{\percent} at \SI{1203}{K}), while Pardo et al.\ additionally reported a second miscibility gap on the S-rich side (from 70 to 95~at.\SI{}{\percent} at \SI{1266}{K}). Furthermore, Spandau and Klanberg \cite{spandau1958thermische} measured the composition and temperature corresponding to the peak of the Ga-rich miscibility gap. Our CALPHAD parameters were optimized to reproduce these invariant reaction points.  

\Cref{fig:Ga-La-S_phase_diagram}(b) shows a section of the Ga--S binary phase diagram focusing on the composition range 48 to 62~at.\SI{}{\percent} and the temperature interval \SI{1100}{K} to \SI{1400}{K}. Lieth et al.~\cite{lieth1966p} and Zavrazhnov et al.~\cite{zavrazhnov2018phase} reported the consistent condition of invariant reactions and liquidus line. Zavrazhnov et al.~\cite{zavrazhnov2018phase} further identified three additional phases ($\sigma$, \ch{Ga2S3}$^\prime$, and $\gamma$-\ch{Ga2S3}) within a narrow composition interval, together with three additional invariant reactions. Our database was optimized to these data and successfully reproduces the phase diagram.

The pseudo-binary \ch{Ga2S3}--\ch{La2S3} phase diagram (\Cref{fig:Ga-La-S_phase_diagram}(c)) was assessed using phase transition data from Refs.~\cite{kamarzin1981growth, loireau1977systemes, bakhtiyarly2016ternary}, including the melting points of \ch{Ga2S3}, \ch{La2S3}, \ch{GaLaS3}, and \ch{La9Ga5S21}, as well as the two eutectic reactions listed in \Cref{tab:Ga-La-S_invariant_reaction}. In the absence of experimental information on the liquidus line, our assessment targeted consistency with the reported invariant points \cite{loireau1977systemes}.

\begin{figure}[H]
  \centering  \includegraphics[width=\linewidth]{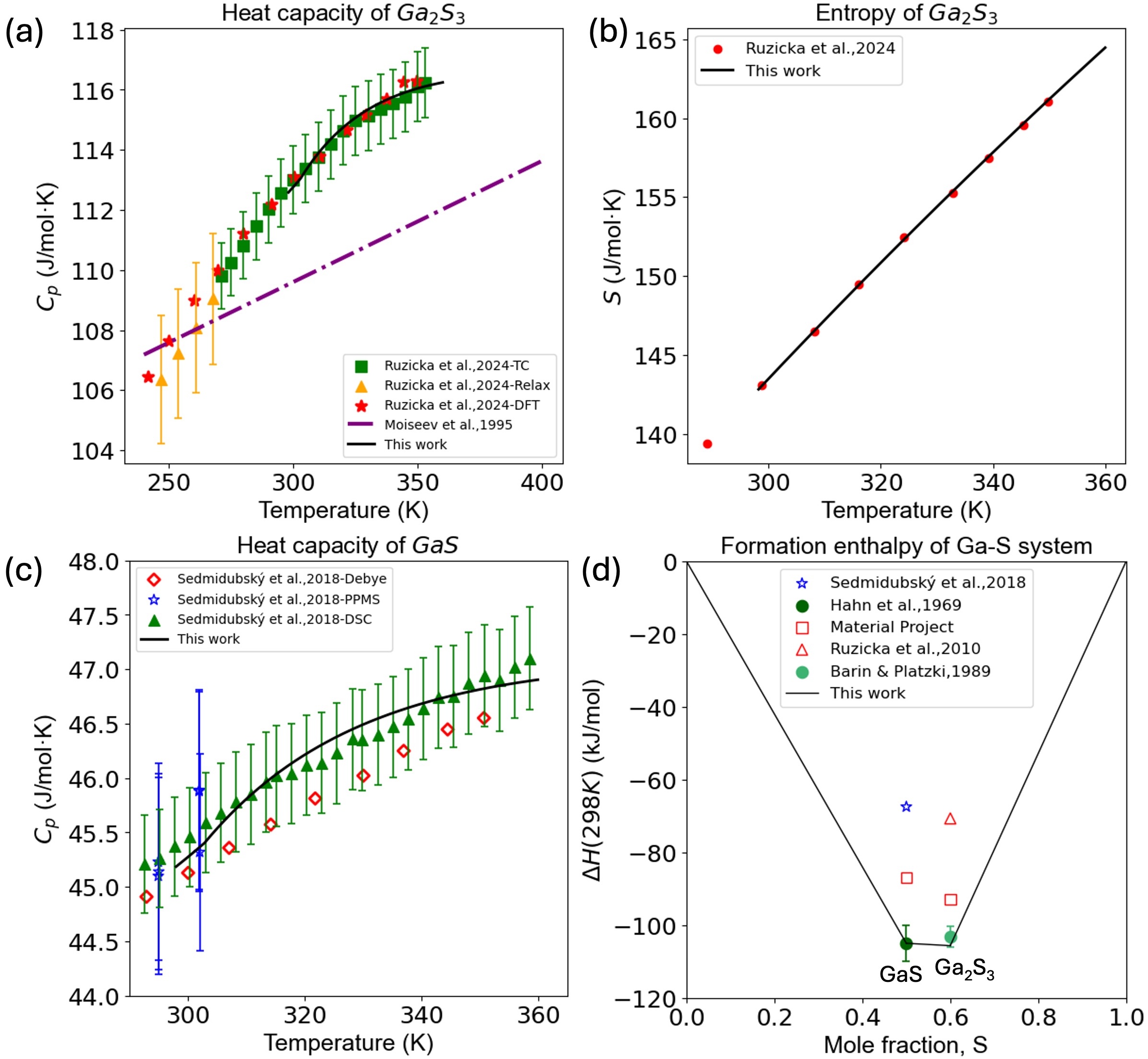}
  \caption{Thermodynamic properties for the Ga--S binary system: (a) heat capacity of \ch{Ga2S3}; (b) entropy of \ch{Ga2S3}; (c) heat capacity of \ch{GaS}; (d) formation enthalpy for the Ga--S system at \SI{298}{K}. Symbols represent literature data: filled for data included in the parameter calibration, open symbols for data shown only for comparison; functional dependencies from literature are represented by lines: ``dashed'' for functions included in calibration, ``dash-dotted'' shown for reference only.}
  \label{fig:Ga-S_thermo_prop}
\end{figure}

\Cref{fig:Ga-S_thermo_prop} summarizes the thermodynamic properties of phases in the Ga--S system. Our assessment adopted the heat capacity and entropy of \ch{Ga2S3} (\Cref{fig:Ga-S_thermo_prop}(a,b)) reported by Růžička et al.\ \cite{ruuvzivcka2024heat}, who combined experimental measurements with DFT calculations. The relaxation calorimetry data carry an measurement uncertainty of approximately \SI{2}{\percent} in the temperature range from 40 to \SI{300}{K}, while the Tian--Calvet calorimeter has a measurement uncertainty of about \SI{1}{\percent} \cite{ruuvzivcka2024heat} (error bars in \Cref{fig:Ga-S_thermo_prop}(a)). For reference, the temperature dependence of the heat capacity estimated using the Neumann--Kopp rule by Moiseev and Sesták \cite{moiseev1995thermochemical} is also included. Deviations in the two sources can be attributed to lattice vibrations, electronic contributions, and anharmonic or defect effects that are not captured by the purely additive assumption of the Neumann–Kopp rule. In addition, our assessment incorporated the heat capacity of \ch{GaS} (\Cref{fig:Ga-S_thermo_prop}(c)) from Sedmidubský et al.\ \cite{sedmidubsky2019chemical}, which is the only source of thermodynamic data for this phase. Their measurements, obtained using PPMS and DSC calorimetry, are reported with measurement uncertainties of approximately \SI{2}{\percent} and \SI{1}{\percent}, respectively \cite{sedmidubsky2019chemical}. Our assessment was primarily calibrated to the DSC results, owing to their broader temperature coverage and lower measurement uncertainty. For comparison, the data obtained from PPMS measurements and the Debye model are also plotted.

\Cref{fig:Ga-S_thermo_prop}(d) presents the formation enthalpies obtained in this study, calibrated to the widely cited experimental values of Hahn et al. \cite{hahn1956uber} for \ch{GaS} and Barin \& Platzki \cite{barin1989thermochemical} for \ch{Ga2S3}. The underestimation of \ch{GaS} formation enthalpy by Sedmidubský et al. \cite{sedmidubsky2019chemical} (DFT) is likely related to the complex stacking sequence of the lamellar \ch{GaS} microstructure, which has a non-trivial influence on the cohesive energy and thus on the calculated formation enthalpy \cite{sedmidubsky2019chemical}. For \ch{Ga2S3}, the discrepancy between the DFT value of Růžička et al.~\cite{ruuvzivcka2024heat} and thermochemical data Barin \& Platzki \cite{barin1989thermochemical} can be attributed to the neglect of finite-temperature contributions in the total energy used for the DFT-based estimation \cite{ruuvzivcka2024heat}. For comparison, \Cref{fig:Ga-S_thermo_prop}(d) also includes the DFT results of Sedmidubský et al.~\cite{sedmidubsky2019chemical} for \ch{GaS}, Růžička et al.~\cite{ruuvzivcka2024heat} for \ch{Ga2S3}, and the calculated values from the Materials Project~\cite{jain2013commentary} for both phases.

\section{Conclusion}

This paper reports the first thermodynamic assessment using the CALPHAD method for binary and pseudo-binary phase diagrams in the Ba--La--S and Ga--La--S systems. Following a thorough literature review, the study chose thermodynamic models appropriate to individual phases and calibrated the thermodynamic parameters to selected experimental and computational data. The assessment incorporated both phase diagram data and thermodynamic properties reported in literature, including the formation enthalpy, entropy, Gibbs energy of formation, and enthalpy increment. 
For liquid phases in the systems that lack reliable data, AIMD calculations were performed to estimate the mixing enthalpy and to complement the limited literature data.
The resulting phase diagrams and calculated thermodynamic properties show good agreement with the available literature data and AIMD calculations in the explored compositional ranges of the binary and pseudo-binary phase diagrams. This study reports an assessment for systems with scarce data, limited to compositional ranges of $0.5 < x_\text{S} < 0.8 $ in the Ba--S, $0.0 < x_\text{S} < 0.7 $ in the La--S, and $0.0 < x_\text{S} < 0.6 $ in the Ga--S systems. The calculated phase diagrams are accordingly most reliable only in these ranges. Outside of these ranges and for liquid phases at high temperatures, thermodynamic predictions 
should be considered provisional due to the lack of reliable experimental data. More detailed and refined CALPHAD models for these systems with strong application potential could greatly benefit from further experimental investigation of high-temperature phase equilibria and thermochemical properties in lanthanide chalcogenide (Ba--Ln--S, Ga--Ln--S) systems.

\section*{Data availability}

The thermodynamic database (TDB) files corresponding to the present assessment are avialable on \href{https://github.com/materials-informatics-az/Chalcogenides-TDB}{GitHub} and as supplementary materials to the \href{https://doi.org/10.1016/j.calphad.2026.102924}{published version of record}. 

\section*{Declaration of competing interest}
The authors declare that they have no known competing financial interests or personal relationships that could have appeared to influence the work reported in this paper.

\section*{Acknowledgments}

This work was supported by Air Force Office of Scientific Research grant No.\ FA9550-23-1-0167. We also thank Dr.\ Rushi Gong for fruitful discussions on the CALPHAD methodology.

\begin{table*}[h]
    \centering
\begin{tabular}{lcccc}
\hline
\multirow{2}{*}{Reaction} & \multicolumn{2}{c}{$T$ (K)} & \multicolumn{2}{c}{Composition $x_S$} \\
\cline{2-5}
 & Literature & Present & Literature & Present \\
\hline
L $\leftrightarrow$ \ch{BaS} & 2454 \cite{boury2021liquid}, 2470 \cite{livey1959high}, 2480 \cite{holcombe1984tentative} & 2475 & 0.5 \cite{boury2021liquid, holcombe1984tentative, livey1959high} & 0.5 \\
({\it{Congruent}}) & (A.~\cite{boury2021liquid}, DTA \cite{holcombe1984tentative, livey1959high}) & & & \\
L+\ch{BaS} $\leftrightarrow$ ht-\ch{BaS2} & 1198 \cite{massalski1990binary} & 1206 & 0.667 \cite{massalski1990binary} & 0.667 \\
({\it{Eutectic}}) & (TA) & & & \\
ht-\ch{BaS2} $\leftrightarrow$ lt-\ch{BaS2} & 973 \cite{massalski1990binary} & 966 & 0.667 \cite{massalski1990binary} & 0.667 \\
({\it{Phase transition}}) & (TA) & & & \\
L+lt-\ch{BaS2} $\leftrightarrow$ \ch{BaS3} & 827 \cite{massalski1990binary} & 831 & 0.75 \cite{massalski1990binary} & 0.75 \\
({\it{Eutectic}}) & (TA) & & & \\
L $\leftrightarrow$ b.c.c.-Ba + f.c.c.-La & 983 \cite{pegaj1985phase} & 983 & 0.004 \cite{pegaj1985phase} & 0.014 \\
({\it{Monotectic}}) & (DTA, XRD) & & & \\
L$\#$2 $\leftrightarrow$ f.c.c.-La + L & 1076 \cite{pegaj1985phase} & 1083 & 0.9 \cite{pegaj1985phase} & 0.9 \\
({\it{Monotectic}}) & (DTA, XRD) & & & \\
L $\leftrightarrow$ \ch{LaS}+b.c.c.-La & 1171 \cite{massalski1990binary} & 1168 & 0.013 \cite{massalski1990binary} & 0.014 \\
({\it{Eutectic}}) & (TA) & & & \\
L $\leftrightarrow$ \ch{LaS} ({\it{Congruent}}) & 2453 \cite{kamarzin1981growth} (TA) & 2473 & 0.5 \cite{kamarzin1981growth} & 0.5 \\
L $\leftrightarrow$ $\gamma$-\ch{La2S3} & 2210 \cite{kamarzin1981growth}, 2133 \cite{nikolaev2008vapor} & 2209 & 0.6 & 0.6 \\
({\it{Congruent}}) & (TA \cite{kamarzin1981growth}, BP \cite{nikolaev2008vapor}, A.~\cite{boury2021liquid}) & & & \\
$\gamma$-\ch{La2S3} $\leftrightarrow$ $\beta$-\ch{La2S3} & 1573 \cite{eyring2002handbook} & 1602 & 0.6 \cite{eyring2002handbook} & 0.6 \\
({\it{Phase transition}}) & (DTA) & & & \\
$\beta$-\ch{La2S3} $\leftrightarrow$ $\alpha$-\ch{La2S3} & 923 \cite{picon1956sous} & 931 & 0.6 \cite{picon1956sous} & 0.6 \\
({\it{Phase transition}}) & (DTA) & & & \\
L $\leftrightarrow$ \ch{La3S4} (\it{Congruent}) & 2383 \cite{kamarzin1981growth} (TA) & 2395 & 0.571 \cite{kamarzin1981growth} & 0.571 \\
L $\leftrightarrow$ \ch{LaS}+\ch{La3S4} ({\it{Eutectic}}) & 1985 \cite{kamarzin1981growth} (TA) & 2005 & 0.544 \cite{kamarzin1981growth} & 0.54 \\
L $\leftrightarrow$ \ch{LaS2} ({\it{Congruent}})  & 1917 \cite{rolland1991vibrational} (TA) & 1921 & 0.667 \cite{rolland1991vibrational} & 0.667 \\
\hline
\multirow{2}{*}{Reaction} & \multicolumn{2}{c}{$T$ (K)} & \multicolumn{2}{c}{Composition $x_{\text{La}_2\text{S}_3}$} \\
\cline{2-5}
 & Literature & Present & Literature & Present \\
\hline
L $\leftrightarrow$ \ch{BaS}+$\gamma$-\ch{La2S3} & 2150 \cite{andreev1991interaction} & 2154 & 0.45 \cite{andreev1991interaction} & 0.41 \\
({\it{Eutectic}}) & (DTA) & & & \\
\hline
\end{tabular}
    \caption{Invariant reactions in the Ba-La-S system. Reaction types and experimental techniques from the literature sources are given in parentheses (A.\ is for thermal arrest, BP is for boiling point, TA is for unspecified thermal analysis).}
    \label{tab:Ba-La-S_invariant_reaction}
\end{table*}

\begin{table*}[h]
    \centering
    \begin{tabular}{cr}
    \hline
        Phase & Calibrated functions and parameters \\
    \hline
        b.c.c.-Ba & $^oG^\text{bcc-Ba}_\text{La} = 10000$\\
         & $^oL^\text{bcc-Ba}_\text{Ba,La} = 34000 - 59.82 T$\\
        f.c.c.-La & $^oL^\text{fcc-La}_\text{Ba} = -5000$\\
         & $^oL^\text{fcc}_\text{Ba,La} = 62000 - 99.71 T$ \\
        b.c.c.-La &$^oG^\text{bcc-La}_\text{Ba} = -1000 - 48 T$\\
        \ch{BaS} & $^oG^\text{BaS}_\text{Ba:S} = ^oG_\text{Ba}^\text{bcc-Ba} + ^oG_\text{S}^\text{Orth}$ \\ 
        & $ - 470318.75 - 88.2 T + 17.29 T \ln(T) - 744851/T + 0.0016534T^2$\\
        ht-\ch{BaS2} & $^oG^{\text{ht-\ch{BaS_2}}}_\text{Ba:S} = ^oG_\text{Ba}^\text{bcc-Ba} + 2^oG_\text{S}^\text{Orth} - 505580 + 7.2T$ \\
        lt-\ch{BaS2} & $^oG^{\text{lt-BaS}_2}_\text{Ba:S} = ^oG_\text{Ba}^\text{bcc-Ba} + 2^oG_\text{S}^\text{Orth} - 507680 - 5  T$ \\
        \ch{BaS3} & $^oG^{\text{BaS}_3}_\text{Ba:S} = ^oG_\text{Ba}^\text{bcc} + 3^oG_\text{S}^\text{Orth} - 551210 + 12 T$ \\
        \ch{LaS} & $^oG^\text{\ch{LaS}}_\text{La:S} = ^oG_\text{La}^\text{dhcp} + ^oG_\text{S}^\text{Orth}$ \\
        & $- 469002.32 - 9.33T + 4.29T\ln(T) - 117987.92/T$ \\
        $\alpha$-\ch{La2S3} & $^oG^{\alpha\text{-\ch{La2S3}}}_\text{La:S} = 2^oG_\text{La}^\text{dhcp} + 3^oG_\text{S}^\text{Orth} - 1226167 + 79 T $\\
        $\beta$-\ch{La2S3} & $^oG^{\beta\text{-\ch{La2S3}}}_\text{La:S} = 2^oG_\text{La}^\text{dhcp} + 3^oG_\text{S}^\text{Orth} $ \\ 
        & $ - 1224293 + 77 T$ \\
        \ch{LaS2} & $^oG^\text{\ch{LaS2}}_\text{La:S} = ^oG_\text{La}^\text{dhcp} + 2^oG_\text{S}^\text{Orth}$ \\ 
        & $-630548.97 + 11T + 0.0077T^2 - 202224.9/T$ \\
        $\gamma$-\ch{La2S3} & $^oG^{\gamma\text{-\ch{La2S3}}}_\text{La:La:Va:S} = 2^oG_\text{La}^\text{dhcp} + 3^oG_\text{S}^\text{Orth}$ \\
        & $- 1177703.12 + 19.78 T + 2.08 T\ln(T)-145168.13/T+0.007935T^2$ \\
          & $^oG^{\gamma\text{-\ch{La2S3}}}_\text{La:La:La:S} = 2.25^oG_\text{La}^\text{dhcp} + 3^oG_\text{S}^\text{Orth}$ 
          \\ 
          & $ - 1241075.91 - 29.92 T + 7.63 T\ln(T) - 309022.23/T+0.0076T^2$ \\
          & $^oL^{\gamma\text{-\ch{La2S3}}}_\text{La:La:La,Va:S} = -2000 + 2 T $ \\
          & $^oL^{\gamma\text{-\ch{La2S3}}}_\text{La:Ba:La:S} = 0.75^oG_\text{Ba}^\text{bcc} + 1.5^oG_\text{La}^\text{dhcp} + 3^oG_\text{S}^\text{Orth} - 1377124 + 115 T $ \\
          & $^oL^{\gamma\text{-\ch{La2S3}}}_\text{La:Ba:Va:S} = 0.75^oG_\text{Ba}^\text{bcc} + 1.25^oG_\text{La}^\text{dhcp} + 3^oG_\text{S}^\text{Orth}$\\
          & $- 1299976 + 110.06 T $ \\
          & $^oL^{\gamma\text{-\ch{La2S3}}}_\text{La:Ba:La,Va:S} = -10000 $ \\
          & $^oL^{\gamma\text{-\ch{La2S3}}}_\text{La:Ba,La:La:S} = -12000 $ \\
          & $^oL^{\gamma\text{-\ch{La2S3}}}_\text{La:Ba,La:Va:S} = -8000 $ \\
    \hline
    \end{tabular}
    \caption{Thermodynamic functions for the solid phases in the Ba--La--S system obtained in this work.}
    \label{tab:Ba-La-S_solid_TDB}
\end{table*}

\begin{table*}[h]
    \centering
    \begin{tabular}{cr}
    \hline
        Phase & Calibrated functions and parameters \\
\hline
        Liquid & $^oL^\text{Liq}_\text{Ba,La} = 30509.54$ \\
        (Ba,La,S,\ch{LaS},\ch{BaS},\ch{La2S3}) & $^1L^\text{Liq}_\text{Ba,La} = 6678.78$ \\
          & $^oL^\text{Liq}_\text{Ba,S} = -300853 $ \\ 
          & $^oG^\text{Liq}_\text{BaS} = ^oG_\text{Ba}^\text{Liq} + ^oG_\text{S}^\text{Liq} - 279853 + 24 T$ \\
          & $^oL^\text{Liq}_\text{BaS,S} = -410000 - 40 T$ \\
          & $^oL^\text{Liq}_\text{Ba,BaS} = -110000 $ \\
          & $^oL^\text{Liq}_\text{La,S} = -344039 $ \\
          & $^oG^\text{Liq}_\text{LaS} = ^oG_\text{La}^\text{Liq} + ^oG_\text{S}^\text{Liq} - 442000 + 44  T$ \\
          & $^oG^\text{Liq}_\text{\ch{La2S3}} = 2^oG_\text{La}^\text{Liq} + 3^oG_\text{S}^\text{Liq} - 867000 + 112 T$ \\
          & $^oL^\text{Liq}_\text{LaS,\ch{La2S3}} = -639000 $ \\
          & $^oL^\text{Liq}_\text{La,\ch{La2S3}} = -440000-25T $ \\
          & $^oL^\text{Liq}_\text{La,LaS} = 20800-25 T$ \\
          & $^1L^\text{Liq}_\text{La,LaS} = 1000$ \\
          & $^oL^\text{Liq}_\text{\ch{La2S3},S} = -240000 - 49 T$ \\
          & $^oL^\text{Liq}_\text{LaS,S} = -240000 - 20 T$ \\
          & $^oL^\text{Liq}_\text{Ba,La,S} = -457411 $ \\
          & $^oL^\text{Liq}_\text{BaS,LaS} = -187391 $ \\
          & $^1L^\text{Liq}_\text{BaS,LaS} = 40000 $ \\
          & $^2L^\text{Liq}_\text{BaS,LaS} = 101739 $ \\
          & $^oL^\text{Liq}_\text{BaS,\ch{La2S3}} = -498386 $ \\
          & $^1L^\text{Liq}_\text{BaS,\ch{La2S3}} = 50000 $ \\
          & $^2L^\text{Liq}_\text{BaS,\ch{La2S3}} = 128385 $ \\
    \hline
    \end{tabular}
    \caption{Thermodynamic functions for the liquid phases in the Ba--La--S system obtained in this work.}
    \label{tab:Ba-La-S_liq_TDB}
\end{table*}

\begin{table*}[h]
    \centering
\begin{tabular}{lcccc}
\hline
\multirow{2}{*}{Reaction} & \multicolumn{2}{c}{$T$ (K)} & \multicolumn{2}{c}{Composition $x_S$} \\
\cline{2-5}
 & Literature & Present & Literature & Present \\
\hline
L $\leftrightarrow$ \ch{GaS} & 1288 \cite{massalski1990binary}, 1235 \cite{lieth1966p} & 1293 & 0.5 \cite{massalski1990binary, lieth1966p} & 0.5 \\ 
({\it{Congruent}}) & (TA, TA) & & & \\
L $\leftrightarrow$ \ch{Ga2S3} & 1393 \cite{massalski1990binary}, 1363 \cite{lieth1966p} & 1390 & 0.6 \cite{massalski1990binary, lieth1966p} & 0.6 \\
({\it{Congruent}}) & (TA, TA) & & & \\
L $\leftrightarrow$ \ch{GaS} + $\sigma$ & 1182 \cite{zavrazhnov2018phase} & 1177 & 0.55 \cite{zavrazhnov2018phase} & 0.55 \\
({\it{Eutectic}}) & (DTA) & & & \\
$\sigma$ $\leftrightarrow$ \ch{Ga2S3}$^\prime$ + L & 1195 \cite{zavrazhnov2018phase} & 1195 & 0.59 \cite{zavrazhnov2018phase} & 0.59 \\
({\it{Peritectic}}) & (DTA) & & & \\
$\sigma$ + \ch{Ga2S3} $\leftrightarrow$ \ch{Ga2S3}$^\prime$ & 1184 \cite{zavrazhnov2018phase} & 1193 & 0.595 \cite{zavrazhnov2018phase} & 0.593 \\
({\it{Peritectic}}) & (DTA) & & & \\
\ch{Ga2S3} $\leftrightarrow$ \ch{Ga2S3}$^\prime$ + L & 1279 \cite{zavrazhnov2018phase} & 1277 & 0.6 \cite{zavrazhnov2018phase} & 0.6 \\
({\it{Peritectic}}) & (DTA) & & & \\
L$\#$2 $\leftrightarrow$ \ch{GaS} + L & 1237 \cite{lieth1966p} & 1232 & 0.47 \cite{lieth1966p} & 0.47 \\
({\it{Monotectic}}) & (TA) & & & \\
L$\#$2 $\leftrightarrow$ \ch{Ga2S3} + L & 1266 \cite{pardo1993diagramme} & 1272 & 0.7 \cite{pardo1993diagramme} & 0.7 \\
({\it{Monotectic}}) & (TA) & & & \\
\hline
\multirow{2}{*}{Reaction} & \multicolumn{2}{c}{$T$ (K)} & \multicolumn{2}{c}{Composition $x_{\text{La}_2\text{S}_3}$} \\
\cline{2-5}
 & Literature & Present & Literature & Present \\
\hline
L $\leftrightarrow$ \ch{Ga2S3}+\ch{LaGaS3} & 1057 \cite{li1999gallium} & 1041 & 0.2 \cite{loireau1977systemes} & 0.21 \\
({\it{Eutectic}}) & (DSC) & & & \\
L+\ch{La9Ga5S21} $\leftrightarrow$ \ch{LaGaS3} & 1225 \cite{loireau1977systemes} & 1215 & 0.5 \cite{loireau1977systemes} & 0.5 \\
({\it{Peritectic}}) & (TA) & & & \\
L $\leftrightarrow$ \ch{La9Ga5S21} & 1425 \cite{loireau1977systemes} & 1417 & 0.643 \cite{loireau1977systemes} & 0.643 \\ 
({\it{Congruent}}) & (TA) & & & \\
L $\leftrightarrow$ \ch{La9Ga5S21}+$\beta$-\ch{La2S3} & 1395 \cite{loireau1977systemes} & 1376 & 0.8 \cite{loireau1977systemes} & 0.78 \\
({\it{Eutectic}}) & (TA) & & & \\

\hline
\end{tabular}
    \caption{Invariant reactions in the Ga-La-S system. Reaction types and experimental methods from literature sources are given in parentheses (TA is for unspecified thermal analysis, DTA is for differential thermal analysis, DSC is for differential scanning calorimetry).}
    \label{tab:Ga-La-S_invariant_reaction}
\end{table*}

\begin{table*}[h]
    \centering
    \begin{tabular}{cr}
    \hline
        Phase & Calibrated functions and parameters \\
    \hline
       \ch{GaS} & $^oG^\text{GaS}_\text{Ga:S} =~^oG_\text{Ga}^\text{Orth} + ~^oG_\text{S}^\text{Orth}$ \\
       & $- 207924.33 - 14.53 T + 4.13 T \ln(T) - 91433.25/T + 0.0026T^2 $\\
       \ch{Ga2S3} & $^oG^\text{Ga$_2$S$_3$}_\text{Ga:S} =~2~^oG_\text{Ga}^\text{Orth} + 3~^oG_\text{S}^\text{Orth}$ \\
       & $ - 512346.89 - 11.98 T + 4.62 T \ln(T) - 209796.95/T + 0.0131T^2$\\
       \ch{Ga2S3}$^\prime$ & $^oG^\text{Ga$_2$S$_3$$^\prime$}_\text{Ga:Ga:S} =~0.41~^oG_\text{Ga}^\text{Orth} + 0.59~^oG_\text{S}^\text{Orth}- 92969 -0.6T$ \\ 
       & $^oG^\text{Ga$_2$S$_3$$^\prime$}_\text{Ga:Va:S} =~0.39~^oG_\text{Ga}^\text{Orth} + 0.59~^oG_\text{S}^\text{Orth}- 91869 + 1.1T $ \\ 
       & $^oG^\text{Ga$_2$S$_3$$^\prime$}_\text{Ga:Ga,Va:S} = -14800+10.3T $ \\ 
       $\sigma$ & $^oG^\sigma_\text{Ga:S} =~0.41~^oG_\text{Ga}^\text{Orth} + 0.59~^oG_\text{S}^\text{Orth} -102634+7.26T$ \\
       \ch{GaLaS3} & $^oG^\text{GaLaS$_3$}_\text{Ga:La:S} =~^oG_\text{Ga}^\text{Orth} 
       +~^oG_\text{La}^\text{dhcp} +3~^oG_\text{S}^\text{Orth} $ \\
       & $- 1107000 + 105 T $\\
       \ch{Ga5La9S21} & $^oG^\text{Ga$_5$La$_9$S$_{21}$}_\text{Ga:La:S} =~5~^oG_\text{Ga}^\text{Orth} + 9~^oG_\text{La}^\text{dhcp} + 21~^oG_\text{S}^\text{Orth}$ \\
       & $- 7996381 + 890 T $\\
    \hline
    \end{tabular}
    \caption{Thermodynamic functions for the solid phases in the Ga--La--S system obtained in this work. The La--S binary parameters are listed in \Cref{tab:Ba-La-S_solid_TDB}. The Ga--La binary parameters are adopted from Ref.~\cite{boudraa2022thermodynamic}. }
    \label{tab:Ga-La-S_solid_TDB}
\end{table*}

\begin{table*}[h]
    \centering
    \begin{tabular}{cr}
    \hline
        Phase & Calibrated functions and parameters \\
    \hline
       Liquid & $^oL^\text{Liq}_\text{Ga,S} = -74639 $ \\
       (Ga,La,S,\ch{GaS},\ch{Ga2S3},\ch{LaS},\ch{La2S3})  & $^oG^\text{Liq}_\text{GaS} = ~^oG_\text{Ga}^\text{Liq} +~^oG_\text{S}^\text{Liq} - 157881$ \\
         & $^oG^\text{Liq}_\text{\ch{Ga2S3}} = 2~^oG_\text{Ga}^\text{Liq} + 3~^oG_\text{S}^\text{Liq} - 371000 + 30 T$ \\
         & $^oG^\text{Liq}_\text{GaS,\ch{Ga2S3}} = -116000 $ \\
         & $^1G^\text{Liq}_\text{GaS,\ch{Ga2S3}} = 60000 $ \\
         & $^oG^\text{Liq}_\text{Ga,GaS} = 35363-8.5T $ \\
         & $^1G^\text{Liq}_\text{Ga,GaS} = 2500 $ \\
         & $^oG^\text{Liq}_\text{\ch{Ga2S3},S} = -45000 $ \\
         & $^1G^\text{Liq}_\text{\ch{Ga2S3},S} = -17000 $ \\
         & $^2G^\text{Liq}_\text{\ch{Ga2S3},S} = 13000 $ \\
         & $^oG^\text{Liq}_\text{GaS,S} = -33000 $ \\
         & $^1G^\text{Liq}_\text{GaS,S} = -38000 $ \\     
         & $^oG^\text{Liq}_\text{GaS,\ch{La2S3}} = -590000 $ \\
         & $^2G^\text{Liq}_\text{GaS,\ch{La2S3}} = -100000 $ \\        
         & $^oG^\text{Liq}_\text{\ch{Ga2S3},LaS} = -410000 $ \\
         & $^1G^\text{Liq}_\text{\ch{Ga2S3},LaS} = 100000 $ \\   
         & $^oG^\text{Liq}_\text{GaS,LaS} = -390000 $ \\
         & $^1G^\text{Liq}_\text{GaS,LaS} = 100000 $ \\
         & $^2G^\text{Liq}_\text{GaS,LaS} = -260000 $ \\
         & $^oG^\text{Liq}_\text{\ch{Ga2S3},\ch{La2S3}} = -780000 - 80 T $ \\
         & $^1G^\text{Liq}_\text{\ch{Ga2S3},\ch{La2S3}} = -20000 $ \\
    \hline
    \end{tabular}
    \caption{Thermodynamic functions for the liquid phases in the Ga--La--S system obtained in this work. The La--S binary parameters are listed in \Cref{tab:Ba-La-S_liq_TDB}. The Ga--La binary parameters are adopted from Ref.~\cite{boudraa2022thermodynamic}.}
    \label{tab:Ga-La-S_liq_TDB}
\end{table*}

\bibliography{refs}
\end{document}